\documentclass[traditabstract,longauth]{aa}

\usepackage{graphicx}

\def\ms{\,m\,s$^{-1}$}         
\def\kms{\,km\,s$^{-1}$}         
\def\ms{\hbox{\,m\,s$^{-1}$}}         
\def\m2s2{\hbox{\,m$^{2}$\,s$^{-2}$}} 
\def\kms{\hbox{\,km\,s$^{-1}$}}       
\def\vsini{\hbox{$v \sin i_*$}}      
\def\Msun{\hbox{$\mathrm{M}_{\odot}$}}             
\def\Rsun{\hbox{$\mathrm{R}_{\odot}$}}
\def\Mjup{\hbox{$\mathrm{M}_{\rm Jup}$}}
\def\Rjup{\hbox{$\mathrm{R}_{\rm Jup}$}}
\def\degr{\hbox{$^\circ$}}
\def\chisq{\mbox{$\chi^2$}}
\def \mp{$M_{\rm p}$}

\newcommand{\halpha}{H$\alpha$} 
\newcommand{\teff}{$T_{\rm eff}$} 
\newcommand{\logg}{$\log g$}

\def \1s{$1\,\sigma$}
\def \kid{$\chi^2$}
\def \t0{T$_0$}
\def \cible{CoRoT-18}
\def \cibleb{{\cible}b}

\begin{document}
\title{Transiting exoplanets from the CoRoT space~mission\thanks{The CoRoT space mission, 
launched on 2006 December 27, has been developed and is operated by CNES, with the 
contribution of Austria, Belgium, Brazil, ESA (RSSD and Science Programme), Germany 
and Spain.}}
\subtitle{XVIII. CoRoT-18b: a massive hot jupiter on a prograde, nearly aligned orbit}

\author{
G.~H\'ebrard\inst{1,2} 
\and T.\,M.~Evans\inst{3} 
\and R.~Alonso\inst{4} 
\and M.~Fridlund\inst{5}
\and A.~Ofir\inst{6}
\and S.~Aigrain\inst{3} 
\and T.~Guillot\inst{7}
\and J.\,M.~Almenara\inst{8,9}
\and M.~Auvergne\inst{10} 
\and A.~Baglin\inst{10}  
\and P.~Barge\inst{8} 
\and A.\,S.~Bonomo\inst{8} 
\and P.~Bord\'e\inst{11} 
\and F.~Bouchy\inst{1,2} 
\and J.~Cabrera\inst{12} 
\and L.~Carone\inst{13} 
\and S.~Carpano\inst{5}
\and C.~Cavarroc\inst{11} 
\and Sz.~Csizmadia\inst{12} 
\and H.\,J.~Deeg\inst{9} 
\and M.~Deleuil\inst{8} 
\and R.\,F.~D\'{\i}az\inst{1,2}
\and R.~Dvorak\inst{14} 
\and A.~Erikson\inst{12}
\and S.~Ferraz-Mello\inst{15} 
\and D.~Gandolfi\inst{5} 
\and N.~Gibson\inst{3} 
\and M.~Gillon\inst{16} 
\and E.~Guenther\inst{17}
\and A.~Hatzes\inst{17} 
\and M.~Havel\inst{7} 
\and L.~Jorda\inst{8} 
\and H.~Lammer\inst{18} 
\and A.~L\'eger\inst{11} 
\and A.~Llebaria\inst{8} 
\and T.~Mazeh\inst{6} 
\and C.~Moutou\inst{8} 
\and M.~Ollivier\inst{11} 
\and H.~Parviainen\inst{9} 
\and M.~P\"atzold\inst{13} 
\and D.~Queloz\inst{4}
\and H.~Rauer\inst{12,19} 
\and D.~Rouan\inst{10}
\and A.~Santerne\inst{8}
\and J.~Schneider\inst{20} 
\and B.~Tingley\inst{9} 
\and G.~Wuchterl\inst{17} 
}

   \offprints{G. H\'ebrard (hebrard@iap.fr)}

\institute{
Institut d'Astrophysique de Paris, UMR7095 CNRS, Universit\'e Pierre \& Marie Curie, 
98bis boulevard Arago, 75014 Paris, France 
\email{hebrard@iap.fr}
\and Observatoire de Haute-Provence, CNRS/OAMP, 04870 Saint-Michel-l'Observatoire, France
\and Department of Physics, Denys Wilkinson Building Keble Road, Oxford, OX1 3RH, UK
\and Observatoire de l'Universit\'e de Gen\`eve, 51 chemin des Maillettes, 1290 Sauverny, Switzerland 
\and Research and Scientific Support Department, European Space Agency,  Keplerlaan1, NL-2200AG, Noordwijk, 
The~Netherlands
\and School of Physics and Astronomy, Raymond and Beverly Sackler Faculty of Exact Sciences, Tel Aviv University, Tel Aviv, Israel  
\and Observatoire de la C\^ote d'Azur, Laboratoire Cassiop\'ee, BP 4229, 06304 Nice Cedex 4, France
\and Laboratoire d'Astrophysique de Marseille, 38 rue Fr\'ed\'eric Joliot-Curie, 13388 Marseille cedex 13, France
\and Instituto de Astrof\'{\i}sica de Canarias, and 
Universidad de La Laguna, Dept. de Astrof\'{\i}sica,
38205 La Laguna, Tenerife,~Spain 
\and LESIA, Observatoire de Paris, Place J. Janssen, 92195 Meudon cedex,~France
\and Institut d'Astrophysique Spatiale, Universit\'e Paris XI, 91405 Orsay, France 
\and Institute of Planetary Research, German Aerospace Center, Rutherfordstrasse 2, 12489 Berlin, Germany
\and Rheinisches Institut f\"ur Umweltforschung an der Universit\"at zu K\"oln, Aachener Strasse 209, 50931, Germany 
\and University of Vienna, Institute of Astronomy, T\"urkenschanzstr. 17, 1180 Vienna, Austria
\and IAG, Universidade de Sao Paulo, Brazil 
\and University of Li\`ege, All\'ee du 6 ao\^ut 17, Sart Tilman, Li\`ege~1,~Belgium
\and Th\"uringer Landessternwarte Tautenburg, Sternwarte 5, 07778 Tautenburg, Germany
\and Space Research Institute, Austrian Academy of Science, Schmiedlstr. 6, 8042 Graz, Austria 
\and Center for Astronomy and Astrophysics, TU Berlin, Hardenbergstr. 36, 10623 Berlin, Germany
\and LUTH, Observatoire de Paris, CNRS, Universit\'e Paris Diderot; 5 place Jules Janssen, 92195 Meudon, France
}
\date{Received ; accepted }

\abstract{We report the detection of CoRoT-18b, a massive hot jupiter transiting in front of its host star with a 
period of $1.9000693 \pm 0.0000028$~days. This planet was discovered thanks to photometric data secured 
with the CoRoT satellite combined with spectroscopic and photometric ground-based follow-up observations. 
The planet has a mass
\mp$\,  = 3.47 \pm 0.38$\,\Mjup, a radius 
$R_{\rm p} = 1.31\pm0.18$\,\Rjup, and a density
$\rho_{\rm p} = 2.2 \pm 0.8$\,g/cm$^3$.
It~orbits a G9V star with a mass 
$M_\star = 0.95\pm0.15$\,M$_\odot$, a radius
$R_\star = 1.00\pm0.13$\,R$_\odot$, and a rotation period 
$P_{\rm rot} = 5.4\pm0.4$\,days. 
The age of the system remains uncertain, with stellar evolution models pointing either 
to a few tens Ma or several Ga, while gyrochronology and lithium abundance
point towards ages of a few hundred Ma.
This mismatch potentially points to a problem in our understanding of the evolution of 
young stars, with possibly significant implications for stellar physics and the 
interpretation of inferred sizes of exoplanets around young stars. 
We detected the Rossiter-McLaughlin anomaly in the \cible\ system thanks to the spectroscopic 
observation of a transit. We measured the obliquity $\psi = 20^{\circ} \pm 20^{\circ}$
(sky-projected value $\lambda = -10^{\circ} \pm 20^{\circ}$), indicating 
that the planet orbits in the same way as the star is rotating and that this prograde 
orbit is nearly aligned with the stellar~equator. 

}

\titlerunning{CoRoT-18b, a massive hot jupiter on a prograde, nearly aligned orbit}
\authorrunning{}

\maketitle

\section{Introduction}

Out of the $\sim550$ exoplanets known to date, more than 100 transit their parent stars as 
seen from the Earth. This particular configuration allows numerous key studies, including 
accurate radius, mass, and thus density measurements 
(see, e.g., Winn~\cite{winn10a} for a review),
atmospheric studies in absorption through transits and in emission through occultations 
(e.g. Vidal-Madjar et al.~\cite{avm03};  Wheatley et al.~\cite{wheatley10}), 
dynamic analyses through possible timing variations (e.g. Holman et al.~\cite{holman10}), 
or spin-orbit alignment measurements thanks to the  Rossiter-McLaughlin effect
(e.g. Bouchy et al.~\cite{bouchy08}). The power 
of these analyses incited numerous search surveys for transiting planets. Most of them were 
discovered in the last five years, and the detection rate is still increasing.

The CoRoT space mission (\textit{COnvection ROtation and planetary Transits}, 
Baglin et al.~\cite{baglin09}) was launched on 2006 December 27. Based on a 
27-cm telescope and a $2.8\degr \times 2.8\degr$-field camera, it is designed to 
study asteroseismology and detect 
transiting exoplanets. The satellite allows several thousand stars ($V = 12 - 16$) to be 
continuously observed for up to 150~days with a high photometric accuracy. CoRoT 
is thus well adapted to detecting transiting planets with small radii, such as CoRoT-7b
(L\'eger et al.~\cite{leger09}; Queloz et al.~\cite{queloz09}), 
or on long orbital periods, such as CoRoT-9b (Deeg et al.~\cite{deeg10}).
It can also detect hot jupiters, such as \cibleb. We report its discovery~here. 

We describe in Sect.~\ref{sect_corot_obs} the CoRoT observations and the 
transit detection of the planetary candidate. 
Then, we present in Sect.~\ref{sect_ground_obs} the 
ground-based follow-up observations that were needed to establish the planetary 
nature of the event detected by CoRoT and also to characterize this planetary system.  
The analysis of the whole dataset and the results are presented in Sect.~\ref{sect_analysis}, 
before conclusion in Sect.~\ref{sect_concle}.

\section{CoRoT observations and transit detection}
\label{sect_corot_obs}

\begin{figure}[t]
 \centering
 \includegraphics[scale=0.51]{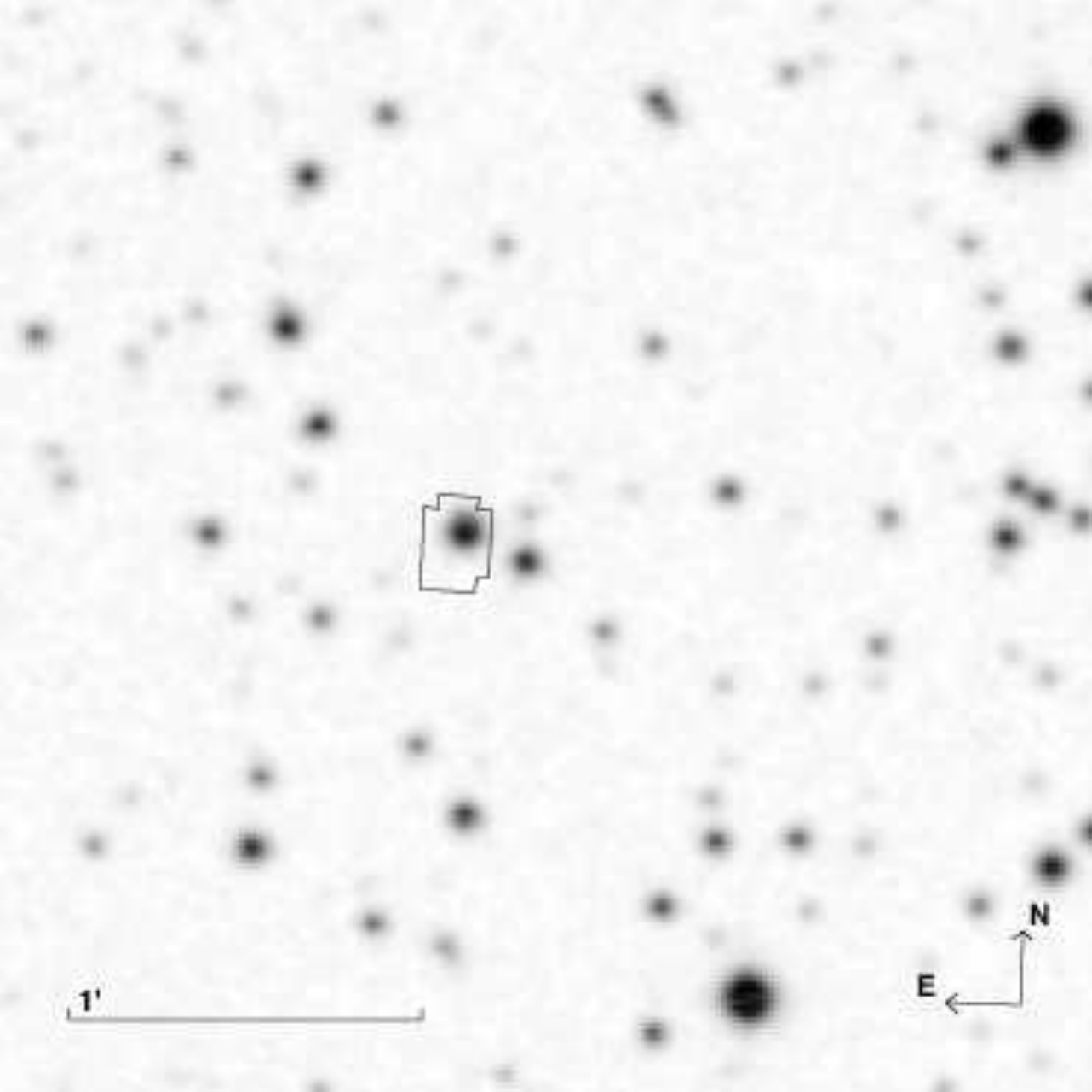}
  \caption{The sky area around \cible\  on the POSS red image. 
The target is in the middle of the image, with the overplot of the 
CoRoT photometric aperture mask. 
    }
  \label{fig_FC}
\end{figure}

\object{\cible} was one of 4161 target stars observed by CoRoT from 2010 March 5 to 29 as 
part of SRa03, the third short run 
of the satellite in the Galactic anti-center direction. The coordinates, magnitudes and identifiers 
of \cible\ in various catalogs are given in Table~\ref{startable}. 
The finding chart is plotted in Fig.~\ref{fig_FC}.
Following the method described in Gandolfi et al.~(\cite{gandolfi08}), the distance, $d$, 
and interstellar extinction, $A_{\mathrm V}$, to \cible\ were derived using the DENIS and 2MASS 
magnitudes reported in Table~\ref{startable} and synthetic colors from a model atmosphere with 
the same parameters as the star (see Sect.~\ref{sect_spectral_analysis}  and 
Table~\ref{starplanet_param_table} below).
We found $A_{\mathrm V}=0.15\pm0.15$~mag and $d=870\pm90$~pc, as well as 
$V=15.00\pm0.10$ in the Johnson standard system.

\begin{figure}[t]
 \centering
  \includegraphics[scale=0.38]{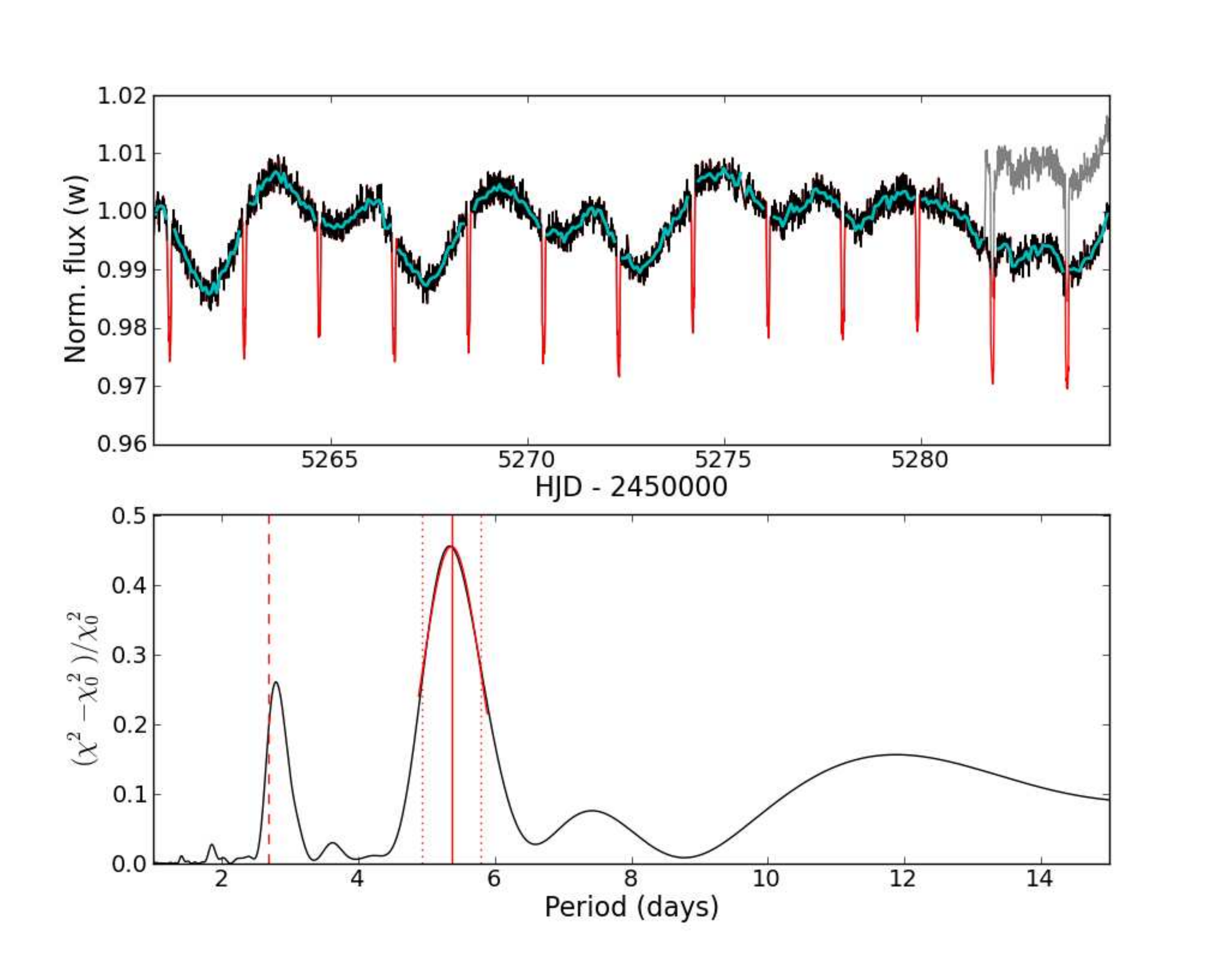}
  \caption{\textit{Top:} CoRoT light curve in black with the 13 transits shown in red.
  All epochs in this paper are given in 
  Heliocentric Julian Date (HJD$_{\rm UTC}$; 	
Eastman et al.~\cite{eastman10}).
  The original light curve (shown in gray) 
  contained a discontinuity at 
  HJD$\,-2450000=5281.62$,
  which was corrected by subtracting 15~mmag from data taken 
  after this date. A linear fit to the light curve has also been used to remove any trend on timescales longer than 
  the duration of the run. 
  The light curve overplotted in blue is binned to one point per orbital period of the CoRoT
  satellite to make sure that no systematics from the rotation period can come through.
  \textit{Bottom:} Lomb-Scargle periodogram of the out-of-transit light curve (black curve 
  in top panel) 
  as a function of relative improvement in \kid\ compared to a constant flux model. It shows  the stellar
  rotation signature at $5.4\pm0.4$~days (solid red line), as well as its first harmonic (dashed red line). 
  Also shown are the Gaussian fit used to estimate the uncertainty in the stellar 
  rotation period and the resulting 1-$\sigma$ interval (dotted red~lines).}
  \label{fig:lc_per}
\end{figure}

\begin{figure}[t]
\centering
\includegraphics[scale=0.455]{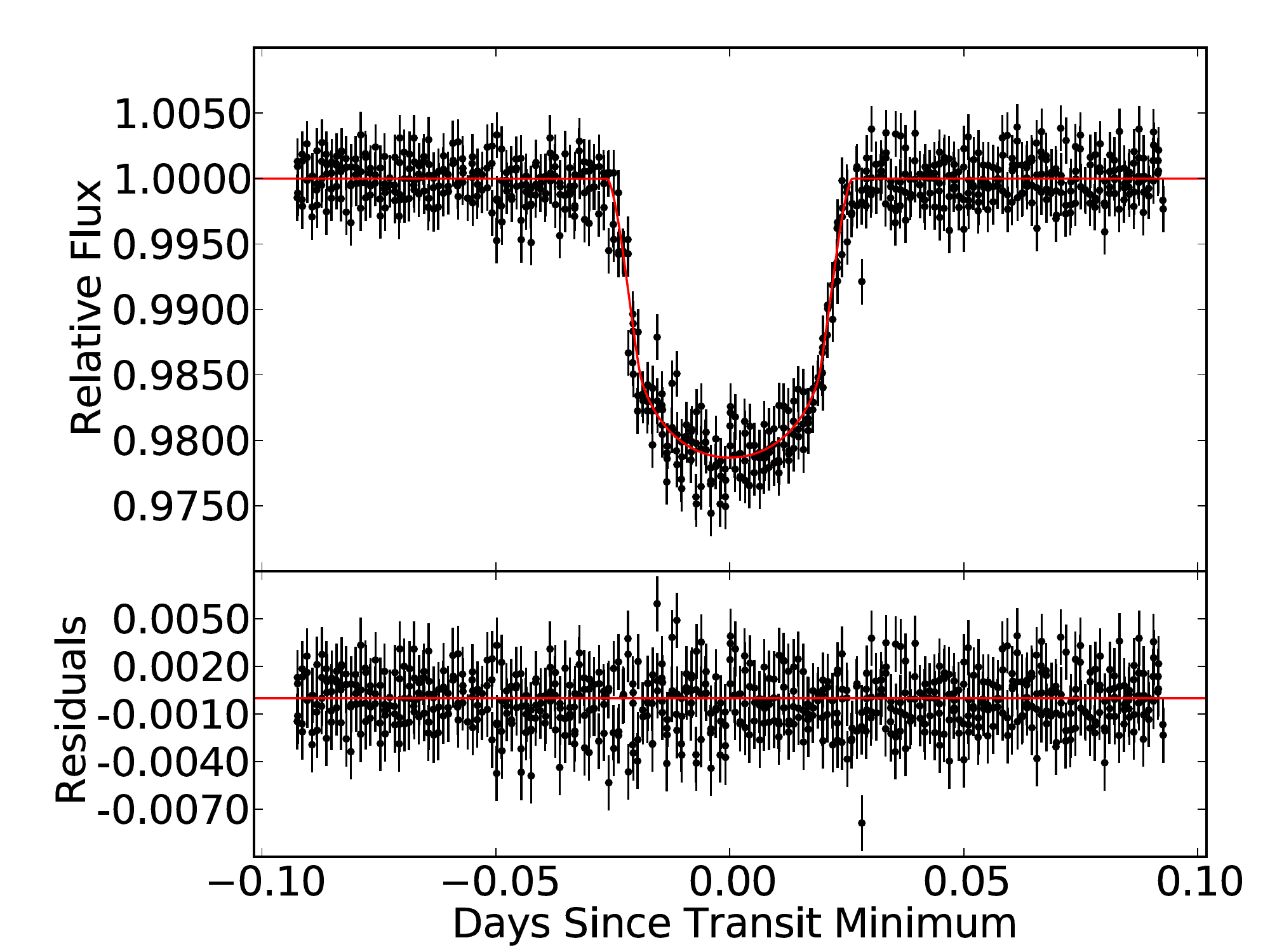}
\includegraphics[scale=0.455]{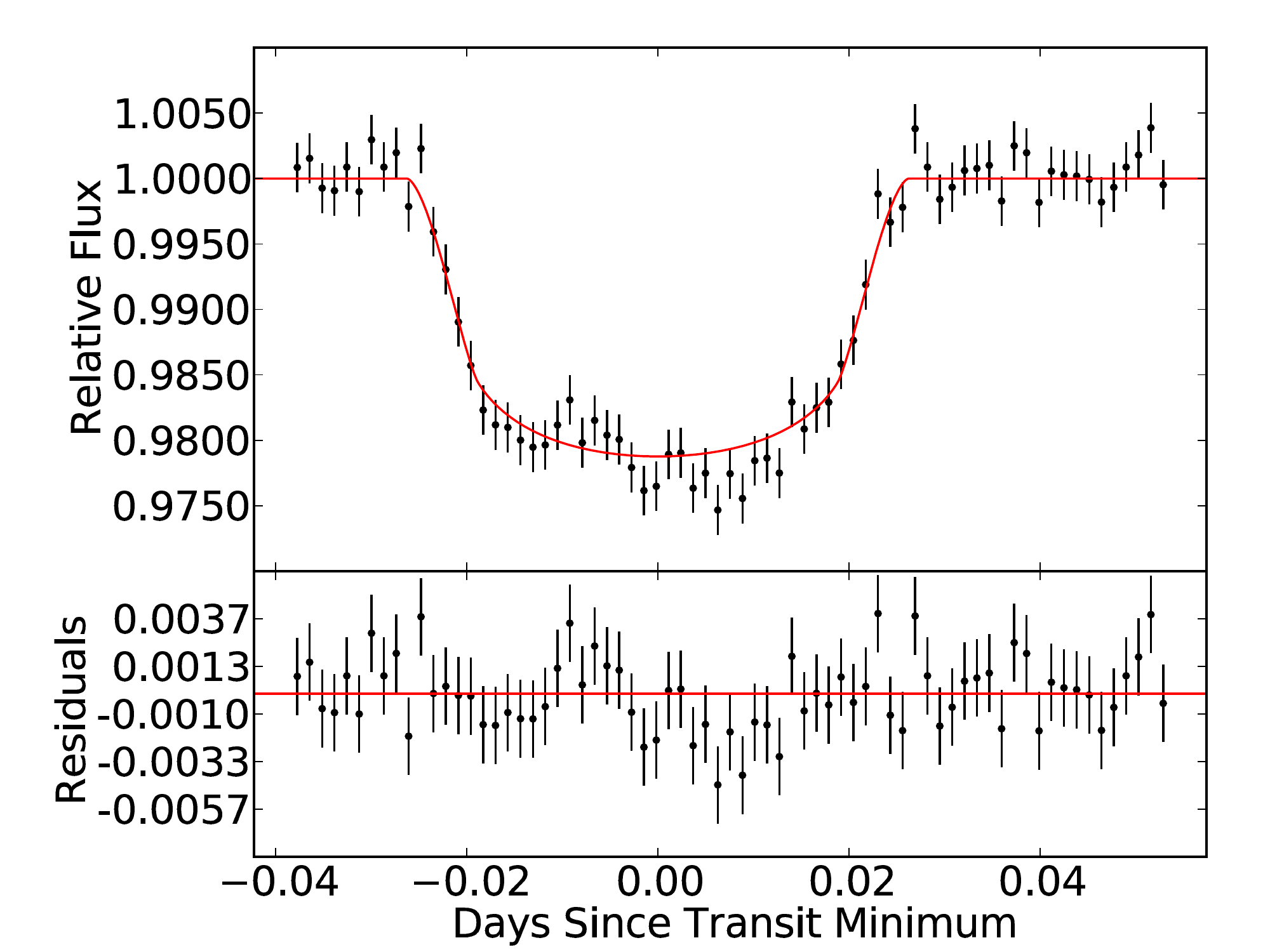}
\caption{Phase-folded CoRoT (top panel) and Euler (bottom panel) transit light curves with the 
best fit and the residuals.
CoRoT data cover 13 transits with a 512-second bin and the Euler data 
one transit with a $\sim3$-minute bin.
Both datasets are joint-fitted (see Sect.~\ref{sect_LC_analysis}).
}
\label{fig_LC_folded}
\end{figure}

The cadence for this target was 32\,seconds throughout the CoRoT observations, resulting in 65120 
exposures spanning the wavelength range 300-1000\,nm. 
The data were processed by the CoRoT pipeline (Auvergne et al.~\cite{auvergne09}). The normalized
 white-light curve, obtained by summing the flux from the three channels and normalizing by the mean flux, 
 is shown in the top panel of Fig.~\ref{fig:lc_per}. 
It clearly shows 13 planetary-like transit features, with a period of $\sim1.9$~days and a depth of $\sim2$\,\%.
The phase-folded light curve is plotted in the upper panel of Fig.~\ref{fig_LC_folded}.

\begin{table}
\caption{\cible\ IDs, coordinates and magnitudes.}
\centering
\renewcommand{\footnoterule}{}
\begin{tabular}{lcc}
\hline\hline
CoRoT window ID & \multicolumn{2}{c}{SRa03\_E2\_1347} \\
CoRoT ID & \multicolumn{2}{c}{315211361} \\
USNO-B1 ID  & \multicolumn{2}{c}{0899-0092144} \\
2MASS ID   & \multicolumn{2}{c}{06324137-0001537}  \\
CMC14 &  \multicolumn{2}{c}{063241.3-000153} \\
\\
\multicolumn{3}{l}{Coordinates} \\
\hline
RA (J2000)  &  \multicolumn{2}{c}{$06^h\,32^m\,41\fs36$} \\
Dec (J2000) &  \multicolumn{2}{c}{$-00\degr\,01\arcmin\,53\farcs71$} \\
\\
\multicolumn{3}{l}{Magnitudes} \\
\hline
\centering
Filter & Mag & Error \\
\hline
B2 (USNO-B1)  & 15.79 & -\\
R2 (USNO-B1)  &14.99 & - \\
V & 15.00 & 0.10 \\ 
R (CMC14) & 14.472 & 0.048  \\
I (DENIS) & 14.051 & 0.030  \\
J (2MASS)  & 13.441 & 0.024\\
H (2MASS)  & 13.080 & 0.031 \\
K (2MASS) & 13.014 & 0.030\\
\hline\hline
\end{tabular}
\label{startable}
\end{table}

To reduce computing time,
the data was rebinned to 512~seconds, which is the normal observing cadence for CoRoT exoplanet 
targets. This binned light curve is used throughout the rest of this paper. 
We checked that this binning does not significantly affect the parameter retrieval
by using the formalism presented by Kipping~(\cite{kipping10b}).
The mean flux was 57366.4\,e$^-$ per 32-second 
exposure, and the relative standard deviation of the binned light curve 
is $6.9\times10^{-3}$, a 
factor 7.2 above the photon noise. This factor reduces to 1.6 when the transits are 
removed and variations on timescales longer than a day are filtered out 
(black line in upper panel of Fig.~\ref{fig:lc_per}; see 
below Sect.~\ref{sect_LC_analysis}).

\cible\ was one of ten objects of interest identified soon after the end of SRa03 observations by 
the ``alarm mode'' pipeline (Surace et al.~\cite{surace08}), which removes the outliers flagged 
by the main pipeline, detrends the light curves to remove long-term trends (instrumental and 
stellar) using a median filter, and searches for transits using an implementation of the box 
least squares (BLS) algorithm of Kov{\` a}cs, Zucker \& Mazeh~(\cite{kovacs02}). 
A~number of tests were then carried out to check that the CoRoT data were compatible 
with a planetary origin for the transits: full transit fits to the white, red, green, and blue 
light curves using the transit formalism of Mandel \& Agol~(\cite{mandel02}), check 
for differences in depth between odd- and even-numbered transits, search for possible 
companion occultation at the transit antiphase, 
and search for ellipsoidal variations. Because~this candidate passed all these tests, it was 
put forward for follow-up observations with high priority.

\section{Ground-based follow-up observations}
\label{sect_ground_obs}

\subsection{``On-off" photometry}

The point spread function of CoRoT that contains 50\%\ of the flux has an elongated 
shape of about $35\arcsec \times 23\arcsec$. The photometry is done through an aperture 
of that size. Owing to this poor spatial resolution, a deep stellar transit diluted in 
the flux of other source(s) also included in the large CoRoT aperture could 
mimic a shallow planetary transit. ``On-off'' photometry of the transit performed 
from the ground with a telescope allowing higher spatial resolution could identify 
contaminated eclipsing binaries (Deeg et al.~\cite{deeg09}).  

``On-off''  photometric observations of \cible\ were performed in November 
2010 with the ESA Optical Ground Station (OGS) 1-m telescope, located at Iza\~na 
in Tenerife (Spain). Alternated short exposures of 30~seconds and long exposures of 100~seconds 
were taken, for a total duration of 28-min observation on-transit and 
28-min observation off-transit. Seven extra sources are detected in a radius of 30\,\arcsec\  
around the main, brighter target.
We performed aperture photometry of the target and neighboring stars. 
The main target shows a transit with a depth $0.03\pm0.01$~mag, in agreement 
with the transit detected with CoRoT. The~seven other targets show stable fluxes, 
within precisions ranging from $0.01$ to $0.39$~mag, depending on the target 
and the exposure time. This ``on-off''  observations thus allowed us to exclude the detected 
transit signature caused by an eclipsing binary diluted in the CoRoT point 
spread~function.

\subsection{Radial velocities}
\label{sect_RV}

We started the spectroscopic follow-up of \cible\ in October 2010 with the 
SOPHIE spectrograph at the 1.93-m telescope of Haute-Provence Observatory, 
France. Three measurements performed in three successive nights near extreme
phases (assuming a circular orbit) showed large radial velocity variations, in phase 
with the CoRoT ephemeris. The variation was on the order of 1~\kms,  indicating 
a companion with a mass around  3\,\Mjup. Thus we decided to pursue the 
spectroscopic observations with SOPHIE to strengthen the detection and 
to characterize the planetary system. We also used the HARPS spectrograph 
at the 3.6-m ESO telescope in La Silla, Chile, and 
the 2.56-m FIES spectrograph attached at the Nordic Optical Telescope in La Palma, Spain. 
Having three ground-based instruments at different longitudes was useful for 
reaching a good phase coverage for this system, which has an orbital period close to 
an integer number of terrestrial days.
The observations were conducted up to January 2011, 
in  good enough weather conditions to allow satisfactory data to be secured in 
reasonable exposure times for this faint target.

Both SOPHIE (Bouchy et al.~\cite{bouchy09}) and HARPS 
(Mayor et al.~\cite{mayor03}) are cross-dispersed, environmentally stabilized 
echelle spectrographs dedicated to high-precision radial velocity measurements.
SOPHIE data were acquired in High-Efficiency mode (resolution power 
$R=40\,000$) and HARPS data in standard HAM mode ($R=115\,000$). 
The spectra extraction was performed using the SOPHIE and HARPS pipelines. 
Following the techniques described by Baranne et al.~(\cite{baranne96}) and  
Pepe et al.~(\cite{pepe02}), the radial velocities were measured from a 
weighted cross-correlation of the spectra with a numerical mask. 
We used a standard G2 mask that includes more than 3500 lines.
The resulting cross-correlation functions were fitted by Gaussians to get the 
radial velocities and the associated photon-noise errors. The full width at half 
maximum of those Gaussians is $12.6 \pm 0.2$~km\,s$^{-1}$, and its contrast is 
$27.7 \pm 0.8$~\%\ of the continuum in the case of the HARPS data. The SOPHIE 
data gave similar parameters.
We adjusted the number of spectral orders used in the cross-correlation in order 
to reduce the dispersion of the measurements. Indeed, some spectral domains  
are noisy, so using them degrades the accuracy of the radial velocity measurement. 
We finally used the orders 10 to 38 for SOPHIE, and 5 to 71 for HARPS.

\begin{figure}[h] 
\begin{center}
\includegraphics[scale=0.515]{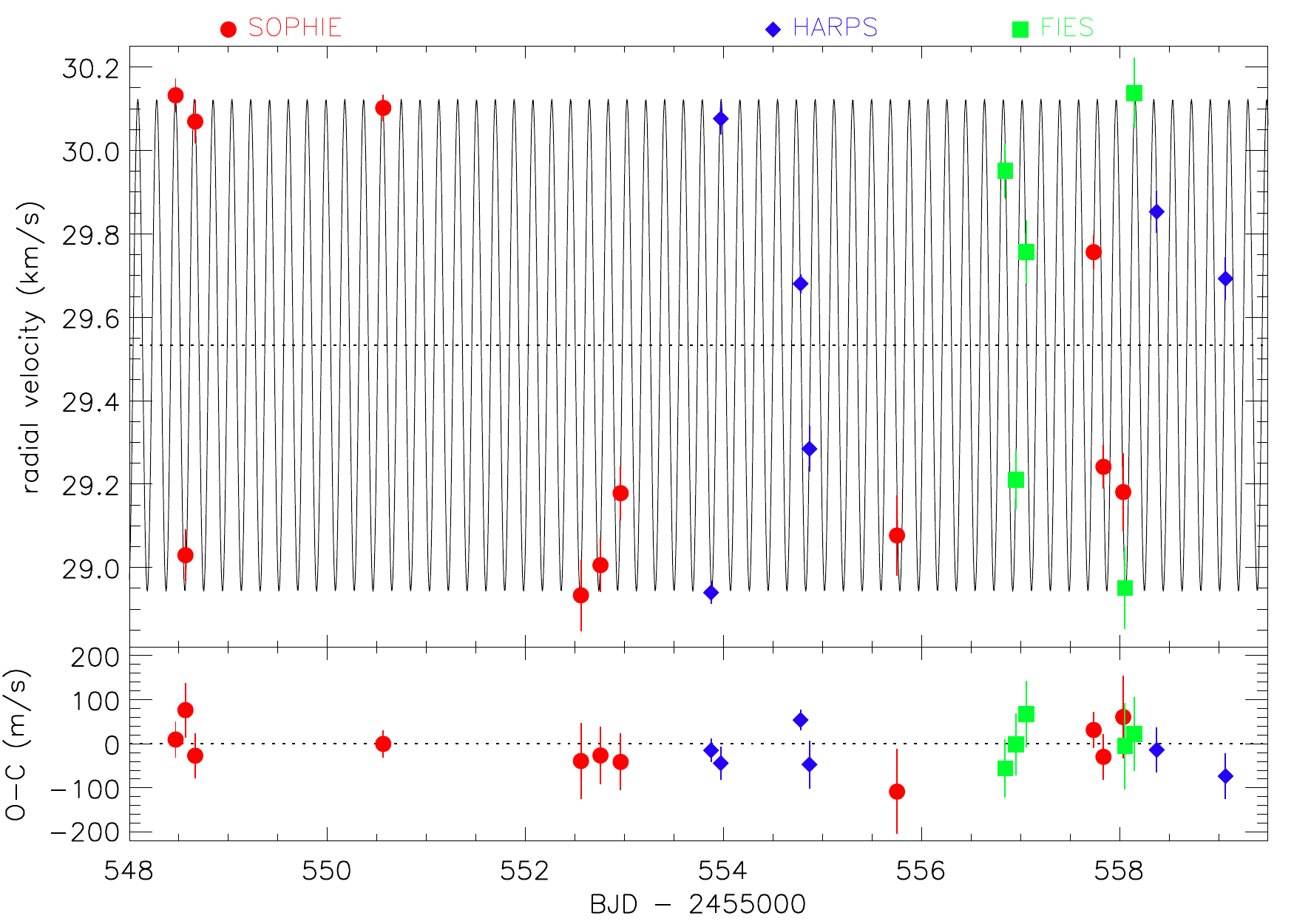}
\includegraphics[scale=0.515]{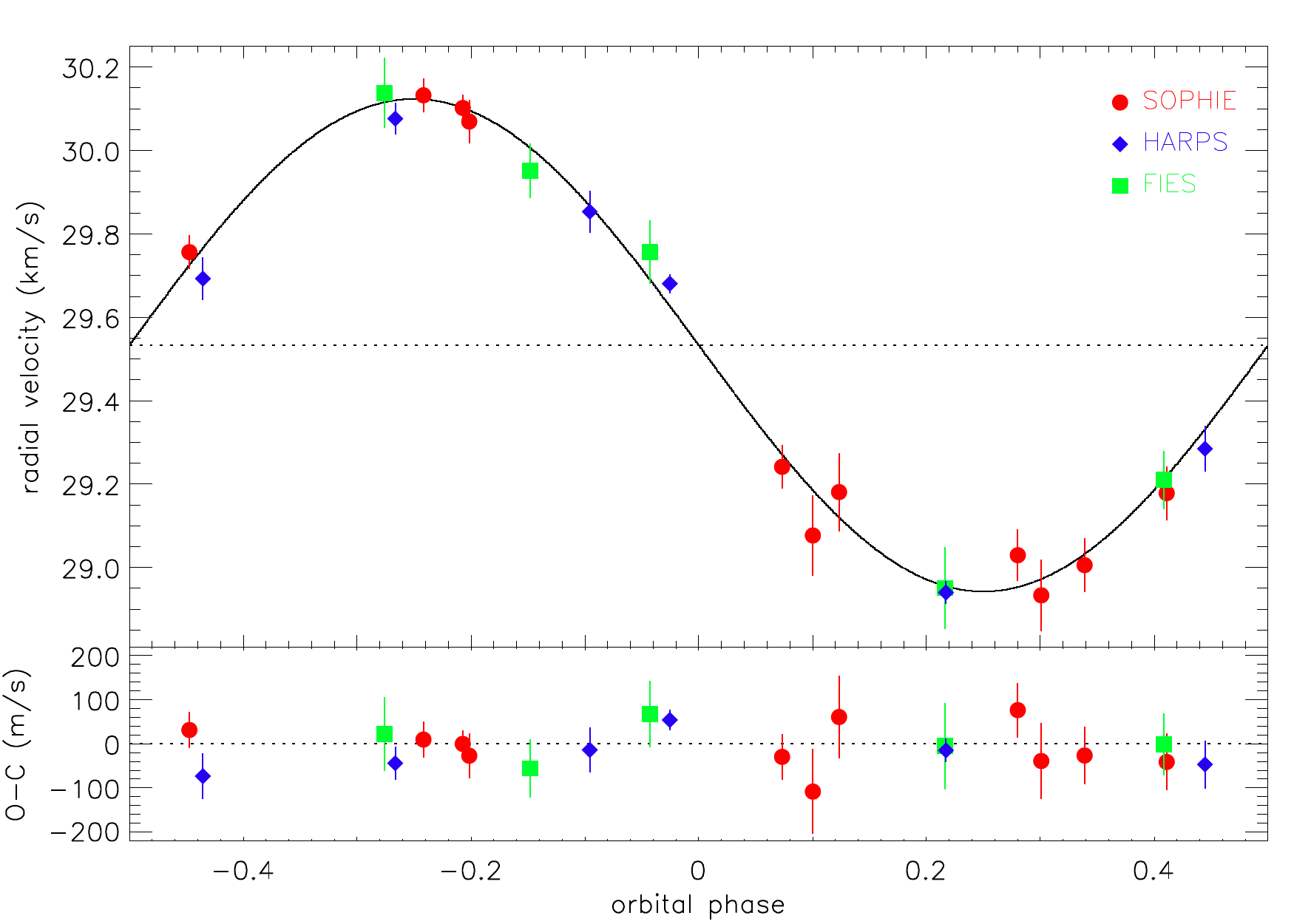}
\caption{\textit{Upper panel:} Radial velocity measurements of \cible\
with 1-$\sigma$\,Êerror bars as a function of time together with their Keplerian 
fit (\textit{top}) and residuals of the fit. 
The fit is described below in Sect.~\ref{sect_RV_fit}. 
\textit{Lower panel:} 
Same as above but as a function of the orbital phase. The data are from 
SOPHIE (red circles), HARPS (blue diamonds), and FIES (green~squares).
}
\label{fig_orb_rv}
\end{center}
\end{figure}

Moonlight contamination was clearly visible in some spectra and in some cases at a radial 
velocity close to that of the target. Such contamination can affect the radial velocity 
measurements. Following the method described in Pollacco et al.~(\cite{pollacco08}) and
H\'ebrard et al.~(\cite{hebrard08}), we estimated and corrected the Moon contamination 
by using the second optical-fiber aperture, which is targeted on the sky for both SOPHIE 
and HARPS, whereas the first aperture points toward the target. This induces radial 
velocity corrections up to $400$\,\ms.

The FIES observations were performed 
in January 2011 under clear and stable weather conditions 
with seeing typically in the range $0\farcs8  - 1\farcs0$. We used the 
$1\farcs3$ high-resolution fiber giving a 
resolving power $R\simeq67\,000$. 
Following the observing strategy described in Buchhave et al.~(\cite{buchhave10}), 
three~consecutive science exposures of 20 minutes  were recorded for each observing 
night immediately preceded and followed by one long-exposed ThAr spectrum. Data 
reduction and spectra extraction were performed using standard IRAF routines. Finally, 
the FIES radial velocity measurements of \cible\ were derived cross-correlating the 
science spectra with the spectrum of the radial velocity standard star HD\,50692 
(Udry et al.~\cite{udry99}) observed with the same instrument~set-up.

The log of the observations and the radial velocity measurements are reported in 
Table~\ref{table_rv}. Radial velocity measurements have accuracies ranging between 
23 and 98\,\ms\ depending on the observing parameters.  
This table also shows the bisector spans that we measured on 
the cross-correlation functions in order to quantify the possible shape~variations 
of the spectral lines. One SOPHIE spectrum was too polluted by the Moon to 
allow any accurate bisector~measurement. 

The radial velocity variations agree with Doppler shifts caused by a planetary 
companion, and the transit-signal detected from the CoRoT light curve
could be interpreted as coming from a massive hot-Jupiter. We designate it as 
\cibleb\ hereafter.

\begin{figure}[h] 
\begin{center}
\includegraphics[scale=0.7]{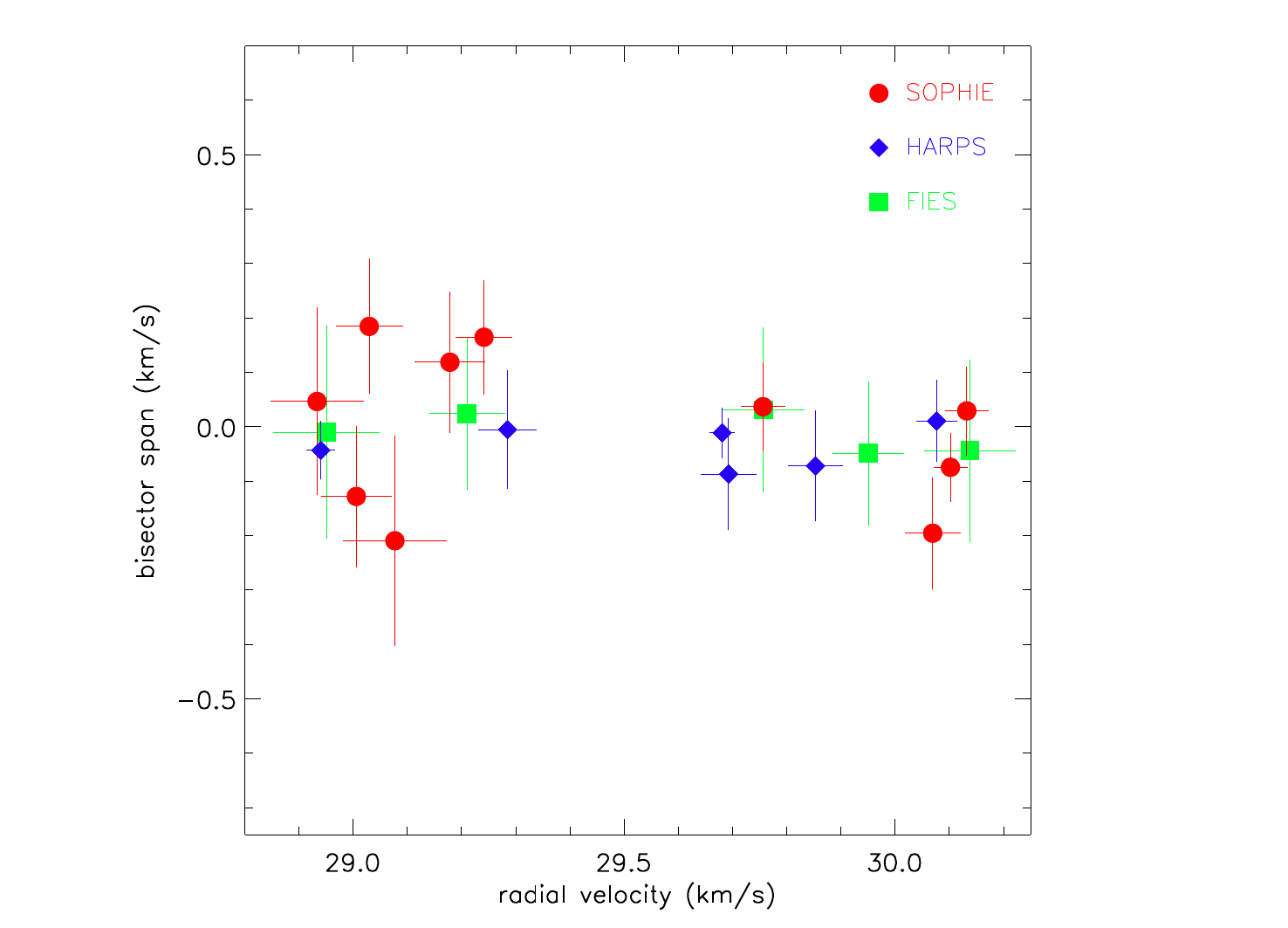}
\caption{Bisector span as a function of the radial velocities with 1-$\sigma$\,Êerror bars. 
The ranges have the same extents in the $x$- and $y$-axes.
The data are from 
SOPHIE (red circles), HARPS (blue diamonds), and FIES (green~squares).}
\label{fig_bis}
\end{center}
\end{figure}

The measurements are displayed in Fig.~\ref{fig_orb_rv}, together with their circular 
Keplerian fit, assuming the period and transit epoch determined by the CoRoT 
light curve and refined with the photometric transit observed from the ground 
(see Sect.~\ref{sect_euler}).
The photometric and radial velocity data show good agreement.
SOPHIE and HARPS radial velocities obtained with different stellar masks (F0 or K5) produce 
variations with the same amplitude as obtained with the G2 mask, so there is no 
indication that their variations could be explained by blend scenarios 
implying stars of different spectral types. Similarly, the cross-correlation function bisector
spans show neither variations nor trend as a function of radial velocity (Fig.~\ref{fig_bis}).
This reinforces the conclusion that the radial velocity variations are not caused by 
spectral line profile changes due to~blends.

\subsection{Transit spectroscopy}
\label{sect_obs_RM}

A transit of \cibleb\ was observed in spectroscopy on 2011 January 28.
The goal was to detect the Rossiter-McLaughlin anomaly, which is an 
apparent distortion of the stellar lines profile because of the transit of the planet in 
front of the rotating star. It allows the measurement of the sky-projected angle 
between the planetary orbital axis and the stellar rotation axis, usually denoted 
$\lambda$ (see, e.g., Bouchy et al.~\cite{bouchy08}). 
The spectroscopic transit was observed with HARPS in the EGGS mode  
to improve the throughput. By comparison to the HAM mode of HARPS
used in Sect.~\ref{sect_RV} for the orbit determination 
($1\arcsec$~diameter fiber with~a~scrambler, allowing the resolution power 
$R=115\,000$ to be reached), the EGGS mode 
of HARPS uses a larger fiber ($1\farcs4$) without scrambler. The spectral 
resolution is then reduced~($R=80\,000$) but the efficiency gain reaches 
a factor $\sim2$~improvement. 

The target was continuously observed during a 5.5-hour sequence under 
good weather condition, with a seeing varying between $0\farcs7$ and 
$1\farcs0$. Twelve measurements were secured with exposure times 
ranging from 1200 to 1800~seconds, including five within the transit.
The remaining observations obtained before and after the transit are 
mandatory for references. The radial velocities were extracted as for 
HARPS/HAM data (Sect.~\ref{sect_RV}), but using fewer orders here for 
the cross-correlation (orders 18 to 69) to reduce the dispersion 
of the measurements.

\begin{figure}[h] 
\begin{center}
\includegraphics[scale=0.525]{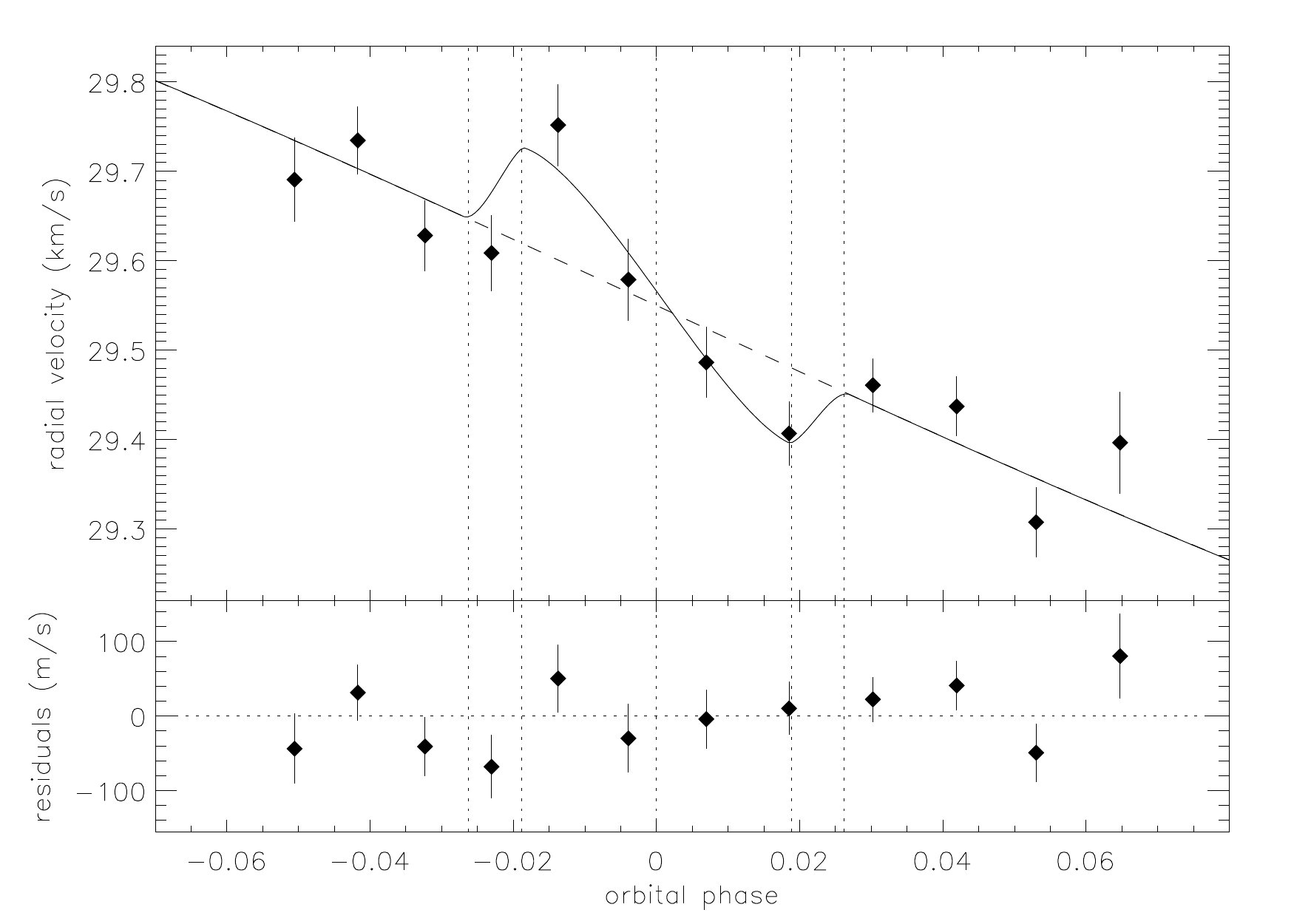}
\caption{Spectroscopic observation of the 2011 January 28 transit of \cibleb. 
\textit{Top:} HARPS/EGGS radial velocity measurements as a function of 
the orbital phase (filled diamonds), Keplerian fit ignoring the transit 
(dashed line), and final fit including the model of the Rossiter-McLaughlin 
anomaly (solid line).
The vertical dotted lines show the times of mid-transit, first, second, third, 
and fourth contacts.  
\textit{Bottom:} Residuals of the final fit.
}
\label{fig_RM}
\end{center}
\end{figure}

The HARPS/EGGS data are plotted in Fig.~\ref{fig_RM}. 
The Rossiter-McLaughlin anomaly is detected, with an amplitude of 
$\sim100$\,\ms, as expected according the rotation of the star and the 
depth of the photometric transit. The red shift during the first part of the 
transit and the blue shift during the second part indicate a prograde 
orbit. The symmetry of the feature agrees with an aligned system.

\begin{table*}
  \centering 
  \caption{Radial velocities of \cible.}
\begin{tabular}{ccccccc}
\hline
\hline
HJD$_{\rm UTC}$ & RV & $\pm$$1\,\sigma$ & Bis. span & exp. time & S/N p. pix. & Instrument/ \\
-2\,455\,000 & (km\,s$^{-1}$) & (km\,s$^{-1}$) & (km\,s$^{-1}$) & (sec) &  (at 550 nm) & mode \\
\hline
484.6704\ \ 			&	30.132	&	0.041	&	0.029	&	3600	&	16	&	SOPHIE/HE \\
485.6621\ \ 			&	29.030	&	0.062	&	0.184	&	3600	&	13	&	SOPHIE/HE \\
486.6469\ \ 			&	30.069	&	0.051	&	-0.196	&	3600	&	11	&	SOPHIE/HE \\
505.6367\ \ 			&	30.102	&	0.032	&	-0.074	&	3600	&	15	&	SOPHIE/HE \\
525.6027$^\dagger$	&	28.933	&	0.086	&	0.046	&	2802	&	8	&	SOPHIE/HE \\
527.5751$^\dagger$	&	29.006	&	0.065	&	-0.128	&	3600	&	11	&	SOPHIE/HE \\
529.6121$^\dagger$	&	29.178	&	0.065	&	0.117	&	3600	&	10	&	SOPHIE/HE \\
557.5227$^\dagger$	&	29.077	&	0.096	&	-0.209	&	3600	&	7	&	SOPHIE/HE \\
577.3831$^\dagger$	&	29.756	&	0.041	&	0.037	&	3600	&	16	&	SOPHIE/HE \\
578.3721$^\dagger$	&	29.241	&	0.052	&	0.165	&	3600	&	15	&	SOPHIE/HE \\
580.3673$^\dagger$	&	29.181	&	0.094	&	-       		&	3600&	13  	&	SOPHIE/HE \\
\hline
538.7441\ \ 			&	28.979	&	0.027	&	-0.043	&	3600	&	11	&	HARPS/HAM \\
539.7253\ \ 			&	30.115	&	0.038	&	0.010	&	3600	&	9	&	HARPS/HAM \\ 
547.7838\ \ 			&	29.720	&	0.023	&	-0.011	&	3600	&	13	&	HARPS/HAM \\
548.6767$^\dagger$	&	29.324	&	0.055	&	-0.005	&	3600	&	7	&	HARPS/HAM \\
583.7515$^\dagger$	&	29.892	&	0.051	&	-0.072	&	3200	&	8	&	HARPS/HAM \\
590.7061\ \ 			&	29.732	&	0.051	&	-0.087	&	3600	&	7	&	HARPS/HAM \\
\hline
568.4512\ \ 			&	30.075	&	0.066	&	-0.049	&	3600&	10	&	FIES \\
569.5086\ \ 			&	29.334	&	0.070	&	 0.024	&	3600&	8	&	FIES \\
570.5516\ \ 			&	29.881	&	0.076	&	 0.031	&	3600&	9	&	FIES \\
580.5441\ \ 			&	29.075	&	0.098	&	-0.010	&	3600&	10	&	FIES \\
581.5089\ \ 			&	30.262	&	0.084  	&	-0.044	&	3600&	11	&	FIES \\
\hline
589.5374 \ \ 			&	29.641	&	0.052	&	0.065	&	1200	&	6	&	HARPS/EGGS \\
589.5542 \ \ 			&	29.719	&	0.042	&	-0.053	&	1500	&	8	&	HARPS/EGGS \\
589.5721\ \ 			&	29.583	&	0.046	&	-0.107	&	1500	&	8	&	HARPS/EGGS \\
589.5897\ \ 			&	29.616	&	0.048	&	-0.037	&	1500	&	7	&	HARPS/EGGS \\
589.6074\ \ 			&	29.727	&	0.051	&	0.046	&	1500	&	7	&	HARPS/EGGS \\
589.6261\ \ 			&	29.596	&	0.052	&	0.156	&	1500	&	7	&	HARPS/EGGS \\
589.6467\ \ 			&	29.501	&	0.045	&	-0.082	&	1800	&	8	&	HARPS/EGGS \\
589.6687\ \ 			&	29.415	&	0.039	&	0.024	&	1800	&	8	&	HARPS/EGGS \\
589.6909\ \ 			&	29.436	&	0.033	&	-0.053	&	1800	&	10	&	HARPS/EGGS \\
589.7132\ \ 			&	29.482	&	0.037	&	-0.168	&	1800	&	9	&	HARPS/EGGS \\
589.7343\ \ 			&	29.329	&	0.043	&	-0.236	&	1800	&	8	&	HARPS/EGGS \\
589.7565\ \ 			&	29.435	&	0.065	&	0.002	&	1800	&	5	&	HARPS/EGGS \\
\hline
\multicolumn{7}{l}{$\dagger$: measurements corrected from Moonlight pollution.} \\
  \label{table_rv}
\end{tabular}
\end{table*}

\subsection{Transit photometry}
\label{sect_euler}
 
A transit of \cibleb\ was observed with the Euler 1.2-m telescope on  
2011 January 28 between 01:08 and 05:36 UT, roughly at the same time 
and location as the Rossiter-McLaughlin anomaly observed in 
spectroscopy with HARPS (see Sect.~\ref{sect_obs_RM}).
The goal was to refine the ephemeris. 
A total of 83 frames were recorded on the recently installed 4K$\times$4K E2V detector, with an 
exposure time of three minutes. 
As for the CoRoT light curve, we checked that this binning does 
not significantly affect the parameter retrieval
by using the formalism presented by Kipping~(\cite{kipping10b}).
Standard calibration images were taken on the same night. 
The fluxes from the target and 20 reference stars were extracted using standard 
aperture photometry with custom IDL routines. An average reference star was constructed 
by interactively  selecting the stars that exhibited less real or instrumental variability. 
Nine stars were selected for this purpose. The final light curve was normalized to the median of 
the flux level after the egress of the transit. It is 
plotted in the lower panel of~Fig.~\ref{fig_LC_folded}.

We considered shot noise as a first estimation 
of the accuracy of the measurements, which was at the level of $10^{-3}$ for this object. 
The accuracy level is later re-established by an evaluation of the reduced \chisq\ of 
the fit process to take the correlated noise present in the data into account 
(see Sect.~\ref{sect_LC_analysis}).
To better estimate  the accuracy of the measurements, the dispersion 
of the data before and after the transit is $1.7\times10^{-3}$. The bump seen on the 
light curve near phase $-0.01$ could hint that the planet is transiting in front of a stellar 
spot; however, the amplitude of this event is within an the order of magnitude of the correlated noise
so a spot detection cannot be claimed here.

\section{Analysis}
\label{sect_analysis}

\subsection{Light curves analysis}
\label{sect_LC_analysis}

\subsubsection{Initial transit fit and light curve normalization}

To remove the modulations in the CoRoT light curve caused by rotating active 
regions on the stellar surface (see the top panel of Fig.~\ref{fig:lc_per}), we fit 
second-order polynomials to stretches of data on either side of each transit spanning 
approximately $-2$ to $-1$ and $+1$ to $+2$ times the transit duration. These 
sections were then normalized and kept for further light curve fitting, while flux 
measurements outside were discarded from the analysis. 

A detailed investigation of the immediate surroundings of the target (using Digital 
Sky Survey data) revealed that $2.0\pm0.1$\% of the flux in the photometric 
aperture was contributed by background stars. This was subtracted from the 
median out-of-transit flux before re-normalizing. The uncertainty in the contamination 
fraction translates into an effective uncertainty on the derived radius ratio due to 
contamination of $\sim0.0001.$ This is 15 times smaller than~the final uncertainties 
derived below, so is~negligible.

The transits were then modeled using the formalism of Mandel \& Agol~(\cite{mandel02}), 
with quadratic limb darkening coefficients $u_1$ and $u_2$ defined according to the 
standard law  of the form
$I(\mu)/I(1)=1 - u_1(1 - \mu) - u_2(1 - \mu)^2$,
where $I(1)$ and $I(\mu)$ are the specific intensities at the center of the stellar disk
and at the angle $\theta$ between the line of sight and the emergent intensity, respectively,
and $\mu=\cos(\theta)$.
We performed an initial least-squares fit (using the Levenberg-Marquardt algorithm)
simultaneously to all 13 CoRoT transits, allowing the following parameters to vary: 
the period $P$, 
the epochs median of transit centers $T_0$, 
the planet-to-star radius ratio $R_{\rm p}/R_\star$, 
the impact parameter $b$,
the scaled semi-major axis $a/R_\star$,
and the combinations of the limb darkening coefficients $u_1+u_2$ and $u_1-u_2$.
We considered a circular orbit. We discuss below the impact of possible low eccentricity 
on our results.
The values for the limb darkening parameters obtained from this analysis were fully
consistent with the values of $u_1=0.47$ and $u_2=0.21$ provided by Sing~(\cite{sing10}) 
for a star with $T_{\rm{eff}}=5500$\,K, $\log g = 4.5$, 
and $[\rm{M}/\rm{H}] = -0.1$ (see Sect.~\ref{sect_spectral_analysis}).
For the remainder of our analysis, we fixed the limb darkening parameters for the 
CoRoT light curve to these values.

\begin{figure}[t]
\centering
\includegraphics[scale=0.65]{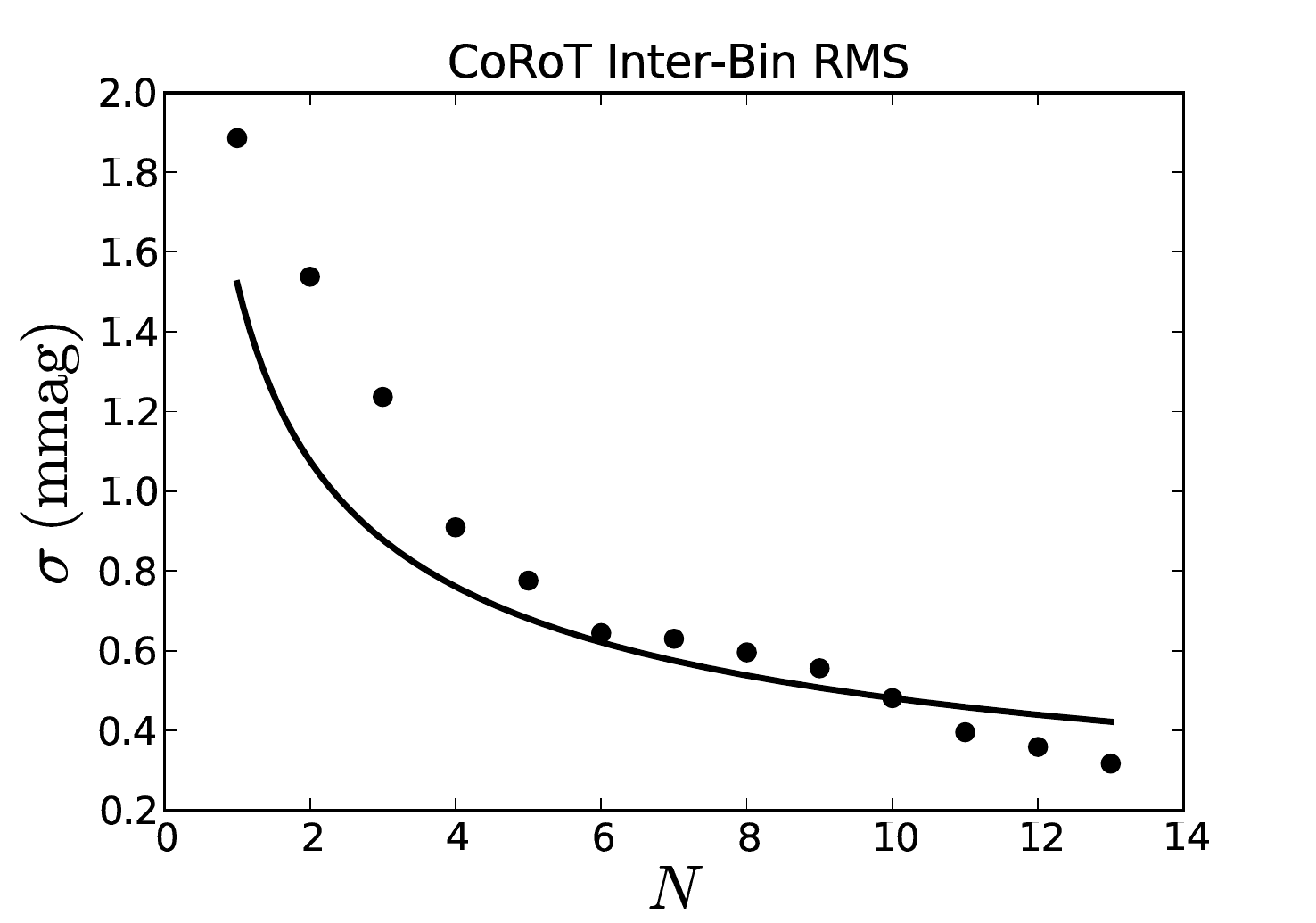}
\includegraphics[scale=0.65]{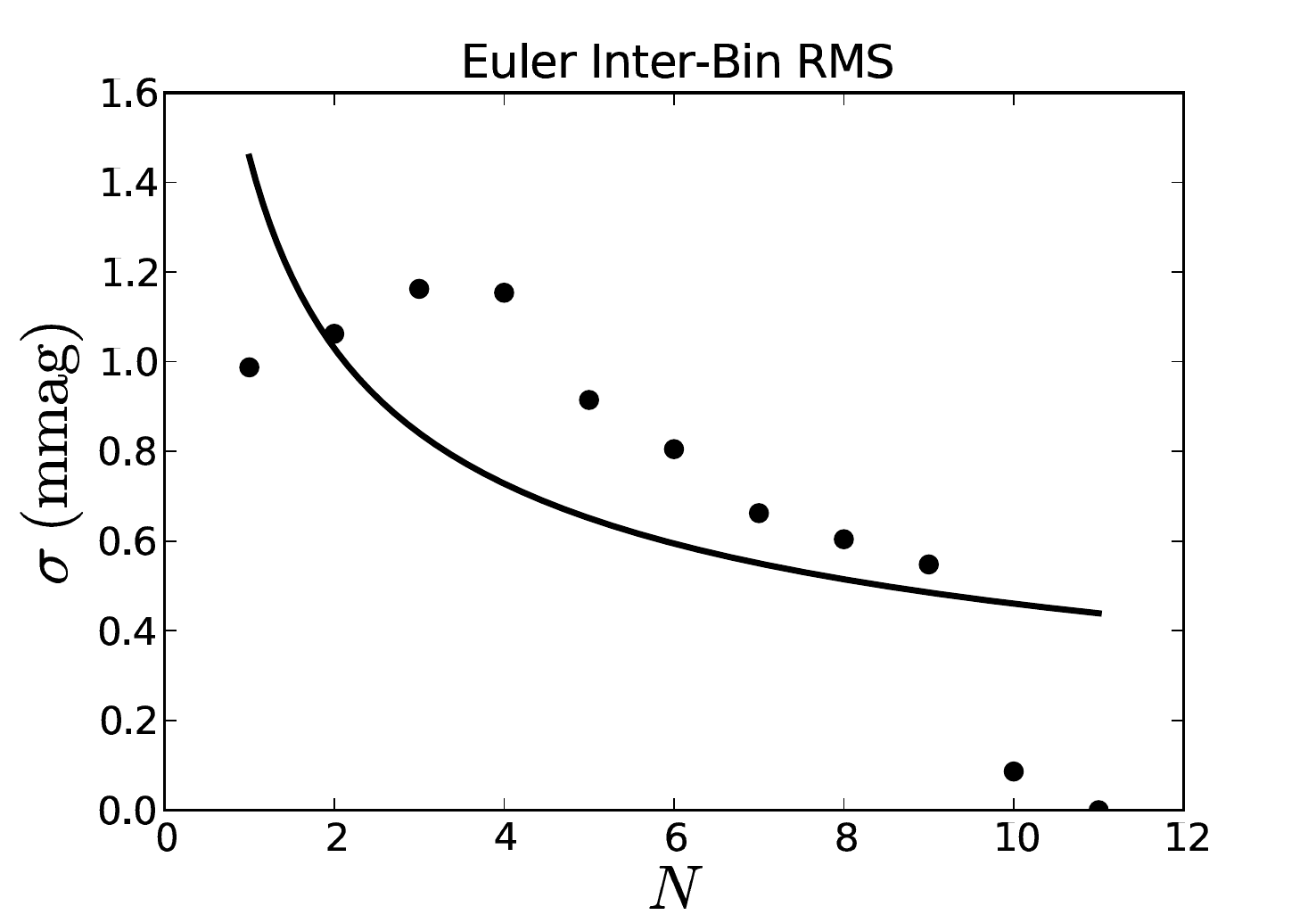}
\caption{The standard deviation $\sigma_N$ of measured out-of-transit flux values as 
a function of the number of points $N$ for the CoRoT light curve (top panel) and the 
Euler light curve (bottom panel). 
The dark curves indicate the best-fit models without red noise, i.e.~only white noise.}
\label{fig:noisemodels}
\end{figure}

\subsubsection{Detailed transit fit}

To estimate the photometric uncertainty of the CoRoT data, we calculated the standard 
deviation of the out-of-transit flux values, except those within 30 minutes of the ingress 
and egress of our initial fit. This provided an estimate of $\sigma=1.7$\,mmag, which 
was then used to perform a 100\,000-step Metropolis-Hastings Markov Chain Monte 
Carlo (MCMC) analysis with the parameters 
$P$, $T_0$, $b$, $a/R_\star$, and $R_{\rm p}/R_\star$ allowed to vary,
 using our initial best-fit values from above as the starting points. We adjusted the 
 jump scales for the free parameters until a step acceptance rate of 25-35\%\ was   
 achieved. The photometric uncertainties were then scaled up to $\sigma=1.9$\,mmag 
 to give a reduced \kid\ of unity for the best-fit MCMC solution. This upwards scaling 
 of the photometric uncertainties can be attributed to the presence of correlated noise 
 in the CoRoT light curve (Pont et al.~\cite{pont06}), often expressed as  
\begin{eqnarray}
\sigma_N^2 &=& \sigma_W^2/N + \sigma_R^2,  
\end{eqnarray}
where $\sigma_N^2$ is the variance between the mean flux values of bins containing 
$N$ data points, and $\sigma_W$ and $\sigma_R$ represent the ``white'' and correlated 
``red'' components of the noise, respectively. The uncertainty on individual flux 
values is then given by $\sigma=\sigma_N$ for $N=1$, so that 
$\sigma=\sqrt{\sigma_W^2+\sigma_R^2}$.  
 The top panel of Fig.~\ref{fig:noisemodels} shows calculated values for $\sigma$ as a function of 
 $N$ for the CoRoT light curve. The variation between the mean flux values of bins 
 containing only a single point ($N=1$) is approximately 1.9\,mmag, equal to the value 
 that produces a reduced \kid\ of unity for the best-fit model. It also shows 
 the expectation for a white-noise-only model that is forced to pass through the point 
 at $N=10$. For lower values of $N$, the white-noise-only model is not able to account 
 for all of the variation in the data. This illustrates why it is necessary to scale the photometric 
 uncertainties upwards from the standard deviation of the out-of-transit flux~values.

A similar procedure was then repeated for the Euler light curve. We first estimated the 
photometric uncertainty to be equal to the standard deviation of the out-of-transit 
measured flux values, which was calculated as $\sigma=1.9$\,mmag. We then 
performed a 100\,000-step MCMC analysis, 
allowing $P$, $T_0$, $b$, $a/R_\star$, and $R_{\rm p}/R_\star$ to vary. 
The limb darkening parameters $u_1$ and $u_2$ 
were also allowed to vary, but this resulted in poor chain convergence. For this 
reason, we decided to set their values to $u_1=40$ and $u_2=30$, as provided  
 by Claret~(\cite{claret04}) for a star with the same parameters as above. To investigate the 
 effect of fixing the limb darkening parameters, we experimented with fixing them 
 to other values provided by Claret~(\cite{claret04}) for stars with similar properties to 
 \cible\ and then performing a least squares fit to the light curve. In all cases, 
 the fitted value for $R_{\rm p}/R_\star$ changed, as expected, while the other parameters 
 remained constant within the uncertainties. In particular, the fitted values for 
 $P$ and $T_0$ were unaffected by varying the limb darkening parameters. This is 
 important because the primary purpose of the Euler light curve is to refine the 
 ephemeris of the orbit (see below). We also verified that the choice of limb 
 darkening parameters for the Euler light curve did not affect the derived values 
 for the other parameters in the simultaneous fitting to the CoRoT and Euler light 
 curves, which is described below. This was done by ensuring that the fitted 
 values for $R_{\rm p}/R_\star$, $a/R_\star$ and $b$ remained the same 
 regardless of whether the Euler light curve was included in the fit.

The bottom panel of Fig.~\ref{fig:noisemodels} shows the variation between binned flux values as a 
function of bin number for the Euler light curve. It shows strongly correlated noise on time scales of 
$\sim10-20$ minutes (between two to five exposures ), which decreases significantly on longer timescales. 
This behavior is poorly modeled by Equation~1 above, with the solid line showing an 
illustrative fit for the white-noise-only case, i.e. $\sigma_R=0$. Instead, we scaled 
the photometric uncertainties up to $\sigma=2.1$\,mmag, which was the value 
required to give a reduced \kid\ of unity for the best-fit~model.

We next performed a joint MCMC analysis on the CoRoT and Euler light curves. 
This was done by initiating five independent chains at random locations in 
parameter space approximately $\sim 5\sigma$ away from the best-fit values 
determined from the initial fitting process. The period was held fixed to the value 
determined from the initial best-fit to the CoRoT light curve, but we allowed the 
transit midtimes to vary for each transit in order to investigate the possibility of 
transit timing variations (see Sect.~\ref{sect_TTV}). 
Values for $a/R_\star$, $R_{\rm p}/R_\star$ and $b$
were allowed to vary with the same values across all transits, and the limb 
darkening coefficients were held fixed to the values described above.

 \begin{figure*}[t]
  \centering
   \includegraphics[width=19.6cm]{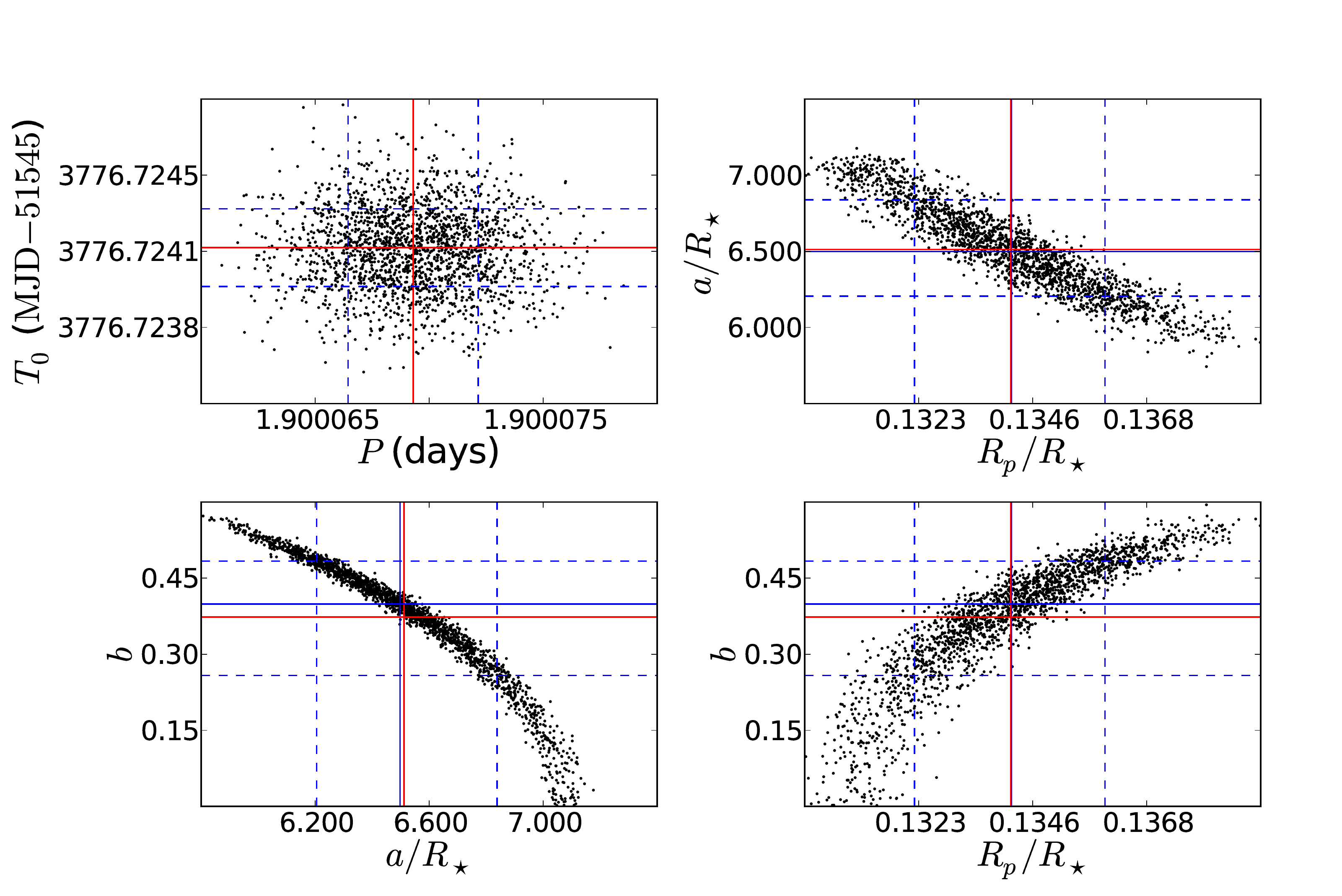}
   \caption{Examples of the results of the MCMC analysis with scatter plots showing the 
   correlations between fitted 
   parameters. Values corresponding 
   to the best-fit (lowest $\chi^2$) solutions are indicated by red lines, while solid blue 
   lines indicate the median values obtained for each parameter, and the dashed blue lines give 
   the upper and lower 1-$\sigma$ uncertainties. The best and median solutions coincide for
   several parameters.}
   \label{fig:scatter}
 \end{figure*}

The five chains were run in parallel for 200\,000 steps. The initial 40\,000 steps of 
each chain were then discarded to allow for the initial burn-in phase as the chains 
settled. We also checked when the \kid\ first dropped below the chain's 
median value, indicating the point at which the local neighborhood of the final 
solution had first been reached. Typically, this occurred within about 20\,000 steps, 
implying that the truncation we made at 40\,000 steps is in fact conservative. The 
Gelman-Rubin statistic was then calculated using the truncated chains
 for each of the free parameters and, in all cases, found to be within approximately 
 1\%\ of unity, indicating good mixing and convergence (Gelman \& Rubin~\cite{gelman92}).              
 The chains were then combined to obtain marginalized posterior distributions for 
 each of the free parameters, with medians and 1-$\,\sigma$ uncertainties reported 
 in Table~\ref{starplanet_param_table} 
 and examples plotted in Fig.~\ref{fig:scatter}. The upper and lower $1\sigma$             
 uncertainties respectively refer to the upper and lower bounds on the intervals 
 containing 34.1\% of the chain steps on either side of the median. 

A refined estimate for the orbital period was then obtained by fitting a model of the form
\begin{eqnarray}
T_0^\prime(n) &=& T_A + n \left(  \frac{T_B-T_A}{173} \right),
\end{eqnarray}
where $T_0^{\prime}(n)$ denotes the transit midtime measured for the $n^\mathrm{th}$ orbit,
$T_A$ is the epoch of the first CoRoT transit and $T_B$ is the epoch of the Euler transit, 
which is 173 orbital periods later. 
Then we ran two 200\,000-step MCMC chains  (one for $T_A$ and one for $T_B$) and  calculated 
the corresponding chain for the estimated period according to $P = (T_B-T_A)/173$.
As expected from P\'al~(\cite{pal09}), the correlation between $T_A$ and $P$ chains was 
negative, while the correlation between the $T_B$ and $P$ chains was positive. We then calculated 
an optimal epoch using Equation~97 of P\'al~(\cite{pal09}), which roughly 
corresponds to the median transit epoch, and find $T_\mathrm{opt} = T_A + 32 \times P$. 
The new chain for this epoch led to the following median value for the optimal epoch, with 
associated 1-$\sigma$ uncertainties: 
$T_0 =T_\mathrm{opt} = 2\,455\,321.72412 \pm 0.00018$~HJD.
By comparison to the classical computation using for example the epoch of the first transit, 
this method with the optimal epoch allows the uncertainty on $T_0$ to be decreased by 2~sec, 
as well as the reduction of the correlation between $T_0$ and $P$ (see Fig.~\ref{fig:scatter}).
The final ephemeris values are reported in Table~\ref{starplanet_param_table}  
and plotted in Fig.~\ref{fig:scatter}.
The best-fit solution for the combined data set has a reduced $\chi^2$ of 1.01.

Finally, we estimated the effect of introducing a small eccentricity in the fit of 
the light curve. Indeed, Kipping~(\cite{kipping10a}) has shown that assuming $e=0$ 
for an eccentric orbit could lead to underestimated uncertainties to 
$a/R_*$ ratio and stellar density $\rho_*$. We derive below 
the upper limit $e<0.08$ at 95\%\ confidence  from the radial velocity measurements 
(see Sect.~\ref{sect_RV_fit}). We performed 
new fits of the CoRoT light curve assuming this extreme eccentricity, with different possible 
values for the longitude $\omega$ of the periastron. Most extreme fits could provide 
$a/R_*$ and  $\rho_*$ values at most to 2\,$\sigma$ lower the values derived above. 
Thus~we slightly decreased the final values on these parameters and increased their 
uncertainties  to account for a small possible~eccentricity.

\subsubsection{Timing analysis}
\label{sect_TTV}

\begin{figure}[ht] 
\begin{center}
\includegraphics[scale=0.26]{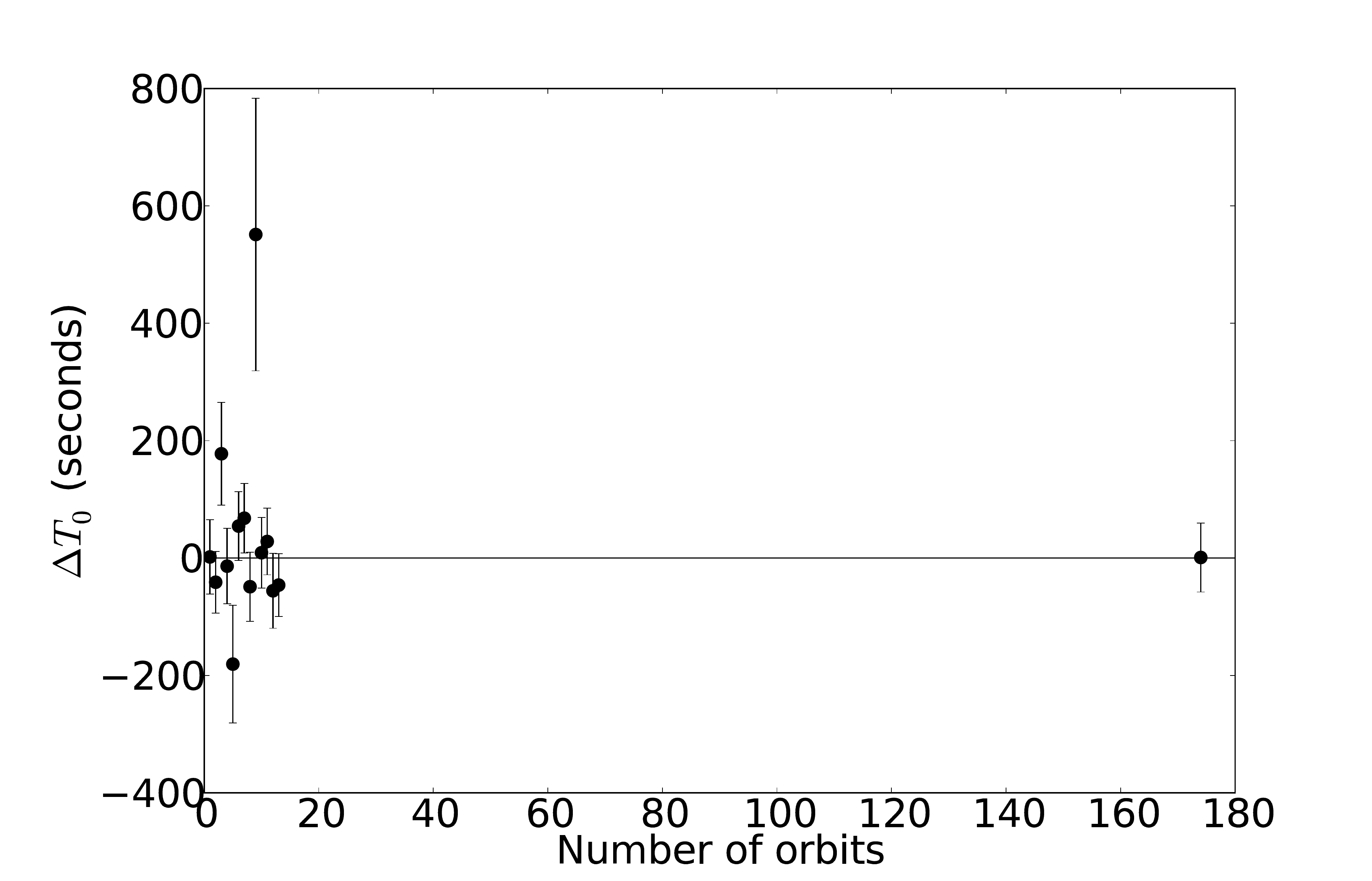}
\caption{Transit timing variation $\Delta T_0$ for each transit of \cibleb. 
The 13 transits observed with CoRoT (on the left) were 
obtained about ten months before those observed with Euler (on the right).
The 5$^{\rm th}$ and 9$^{\rm th}$ CoRoT transits do cover neither the ingress nor egress
and therefore could conduct to erroneous timing measurements.}
\label{fig_TTV}
\end{center}
\end{figure}

The 13 successive transits observed by CoRoT and the observation of the additional transit with 
Euler offer an opportunity for transit timing variations (TTVs) research~(Holman \& Murray~\cite{holman05}). 
Figure~\ref{fig_TTV} shows the TTVs measured~by
\begin{eqnarray}
\Delta T_0(n) = T_0^{\prime}(n) - T_0,
\end{eqnarray}
where $T_0^{\prime}(n)$ again denotes the transit midtime measured for the $n^\mathrm{th}$ orbit 
and $T_0$ denotes the refined transit midtime derived by fitting to the ephemeris equation, 
as described above. We obtain a reduced $\chi^2$ of 1.3 for the hypothesis that there are 
no TTVs, i.e. $\Delta T_0(n)=0$ for $n=1,\ldots,14$. When the ninth transit is removed 
(visible as the most discrepant point in Fig.~\ref{fig_TTV}) the reduce $\chi^2$ improves to 0.98. 
Inspection of this particular transit reveals that neither the ingress nor the egress was sampled, 
which could help explain why the measured TTV is somewhat discrepant.  
This is also the case for the fifth transit. Also, it is possible 
that the uncertainties in the measured transit midtimes are slightly underestimated because 
we did not account for unocculted spots that could result in different values for the 
effective radius ratio $R_p/R_\star$ being measured from transit-to-transit as the spot coverage 
evolves; instead, we held $R_p/R_\star$ fixed for all transits in the MCMC analysis. In either case, 
given the perfect agreement of the Euler ephemeris with all of the other CoRoT ephemerides, 
we conclude that there is no evidence of TTVs in the combined data set. Given that the 
characteristic uncertainty on our TTV measurements was about 60~seconds, we obtain a 
3-$\sigma$ TTV upper limit of~180~seconds. 

\subsubsection{Rotation period}
\label{sect_rota_period}

The CoRoT white-light  curve of \cible\, shown in the top panel of Fig.~\ref{fig:lc_per} 
displays obvious signs of rotational modulation of star spots, with at least two large 
active regions present on the stellar surface at any one time. To estimate the rotation period 
we first cut out a section lasting twice the transit duration around each transit, then 
fit a straight line to the out-of-transit light curve to remove any long-term trend. We 
also corrected for a discontinuity near the end of the light curve by subtracting 
15\,mmag from all data taken after the discontinuity. 
This detrending procedure is slightly different from the one presented above in 
Sect.~\ref{sect_LC_analysis}.

We estimated the rotation period by fitting sinusoids at 1000 trial periods ranging 
from 1 to 15 days to the corrected out-of-transit data. The resulting best-fit 
amplitudes are shown as a function of trial period in the bottom panel of 
Fig.~\ref{fig:lc_per}. The dominant peak clearly corresponds to the interval 
between repeat appearances of individual active regions, hence to the 
rotation period. There is also significant power at the first harmonic of the 
rotation period. 
We do not expect aliases here from the CoRoT satellite orbital period, because it is well 
below the Nyquist sampling frequency of the data. We checked for that by binning 
the light curve to one point per orbital period (upper panel of Fig.~\ref{fig:lc_per}) to 
make sure that no systematics from the satellite  rotation period can come through, 
and repeated the study. The results were identical.

To refine the estimate of the rotation period, we fit a Gaussian 
to the periodogram around the main peak, shown in the lower panel of Fig.~\ref{fig:lc_per}, 
and adopted the standard deviation of the Gaussian as our period uncertainty, 
yielding $P_{\rm rot} = 5.4 \pm 0.4$ days. This period is quite short for a late G-type 
star: even at the age of the Hyades, typical rotation periods for this spectral type are
 in the range of eight to nine days (Kawaler~\cite{kawaler89}).

\subsubsection{Planetary occultation upper limit}

We derived the depth  upper limit of a possible occultation of
the planet by the star by fitting a Mandel \& Agol~(\cite{mandel02}) transit
model at phase 0.5, with no limb darkening and a transit depth reduced by a 
factor $F_p/F_*$,  where $F_p$  is the planetary flux and $F_*$ the stellar flux. 
All parameters were kept
fixed to the best-fit values derived from the transit, except for the
eclipse depth, for which we obtain a best-fit value of 
$0.05 \pm 0.19$\,mmag, i.e. consistent with zero. The resulting 
3-$\sigma$ upper limit is 0.61\,mmag. 
This is not particularly stringent, as one can expect a planetary to star
flux ratio of $0.40$\,mmag in the optimistic case of a geometric albedo 
equal to one.

\subsection{Stellar analysis and classification}
\label{sect_spectral_analysis}

To determine the planetary parameters with an as high as possible precision 
we need to know the physical conditions of its host star. 
Seager \& Mall\'en-Ornelas~(\cite{seager03}), among others, have shown that ideally 
there is one stellar parameter, the
 stellar density, which can be obtained from a transit light curve of sufficient photometric 
 precision. From this parameter, it is possible (with a number of assumptions) to derive, 
 through modeling, other physical parameters of the system. Nevertheless, as pointed 
 out in this context by e.g. Fridlund et al.~(\cite{fridlund10}), high-precision photometric and
 spectroscopic measurements that has been carried out on other exoplanet host
 stars, do suggest that this rarely infers reliably to the other properties of
 the star -- mainly because of flaws in stellar theory (Winn et al.~\cite{winn08}).
 
 We used the two sets of HARPS observations to perform this analysis: 
the HARPS/HAM data  (co-addition of six spectra 
 totaling 5.8~hours of integration,  see Sect~\ref{sect_RV}) and  
 the HARPS/EGGS data (co-addition of 12 spectra 
 totaling 5.4~hours of integration,  see Sect~\ref{sect_RV}). Due to its lower resolution power ($80\,000$ 
 \textit{vs.} $115\,000$) the HARPS/EGGS  data presents the higher signal-to-noise 
ratio ($\sim35$ in the continuum at \halpha).
 We analyzed both sets of 
 observations and find no significant differences in the stellar parameters beyond 
 the internal 1-$\sigma$ error. Three observations are immediately made while 
 inspecting the co-added spectra. The appearance is that of a cool star, the line
  profiles are relatively broad (\vsini$=8.0\pm1.0$\,\kms), and there is a faint absorption 
  (equivalent width of $\sim40$\,m\AA) at the location of the 
 \ion{Li}{i} (6707.8 \AA) line. 
 There is no obvious detection of any \ion{Ca}{ii} chromospheric emission.
This \vsini\ direct measurement agrees with this derived from the cross-correlation 
function following the Santos et al.~(\cite{santos02}) methodology.

To determine the spectroscopic parameters, we used the 
Spectroscopy Made Easy (SME, version 162, February 2011) software package
(Valenti \& Piskunov~\cite{vp96}; Valenti \& Fischer~\cite{vf05}). 
SME calculates synthetic spectra and fits the observations to it. All the normal stellar parameters 
(\teff, \logg, \vsini, $[\rm{Fe/H}]$, abundances, etc.) 
can be used either as input or as free parameters to solve for. A grid of stellar models (Kurucz models)
was utilized in order to determine the fundamental
stellar parameters iteratively. This was achieved by fitting the observed spectrum
directly to the synthesized spectrum and minimizing the discrepancies
using a nonlinear least-squares algorithm. SME requires atomic line data 
in order to synthesize   a spectrum. We utilized input from the Vienna Atomic Line Database 
(Kupka et al.~\cite{kupka99}; Piskunov et~al.~\cite{piskunov95}).

Using SME and a sample of more than 1000 stars, Valenti \& Fischer~(\cite{vf05}) found uncertainties 
 of 44\,K in \teff, 0.06~dex in  \logg, and 0.03~dex in [M/H] per measurement.  
 Based on the CoRoT material (stars hosting CoRoT 
 planets or candidates that have not  been positively shown to host planets),
  we find slightly larger errors than  Valenti \& Fischer~(\cite{vf05}):
  $70-100$\,K in \teff, depending on the signal-to-noise in the continuum of the spectrum at the location of the 
  Balmer lines, $0.05-0.1$\,dex in \logg, depending also on the spectral type 
  and  on which ion we used, and finally 
$0.05-0.1$\,dex in [M/H].  However, by comparing the measurements with model 
isochrones they found a larger, systematic offset of $\sim0.1$\,dex and a scatter 
that can occasionally reach $0.3$\,dex in \logg. In \cible, we find an internal 
discrepancy using SME of $0.1$\,dex. We therefore assign $0.1$\,dex as our 1-$\sigma$~precision. 

We found \teff\,$= 5443K\pm100$\,K from the profile of the Balmer lines. 
We determined the metallicity and found consistent results from different ions indicating a star 
of slightly lower than Solar metallicity: $[{\rm M/H}]\simeq-0.1\pm0.1$. 
The \logg\ was determined utilizing the lines of \ion{Ca}{i}, \ion{Mg}{i} and \ion{Na}{i}, finding a consistent result 
 of $4.4\pm0.1$.  The spectroscopic parameters of \cible\  
are summarized in Table~\ref{starplanet_param_table}.

\subsection{Stellar evolution tracks and the age problem}
\label{sect_age}

Altogether, \cible\ seems remarkably similar to CoRoT-2 (Alonso et al.~\cite{alonso08}): 
the two stars (\cible\ \textit{vs.} CoRoT-2) have comparable effective temperatures (5440 \textit{vs.} 5450~K), 
metallicities ($-0.1$ \textit{vs.} $0.0$), spin periods (5.4 \textit{vs.} 4.5~days), 
and \vsini\ (8.0 \textit{vs.}  12\,\kms), and they are both active, 
with peak-to-peak photometric variabilities of $\sim 2\%$ and $\sim 4\%$, respectively. 
In addition 
\cibleb\ and CoRoT-2b are the only known planets (transiting or not) 
orbiting a star colder than 6000~K and with a large \vsini. The other 
planet-host stars in this temperature range all have \vsini\ values 
in the range $[0-5]$\,\kms. All the planet-host stars having 
\vsini$\;\sim10$\,\kms\ or larger are F-stars -- except \cible\ and CoRoT-2.

\begin{figure}[h] 
\begin{center}
\includegraphics[scale=0.26]{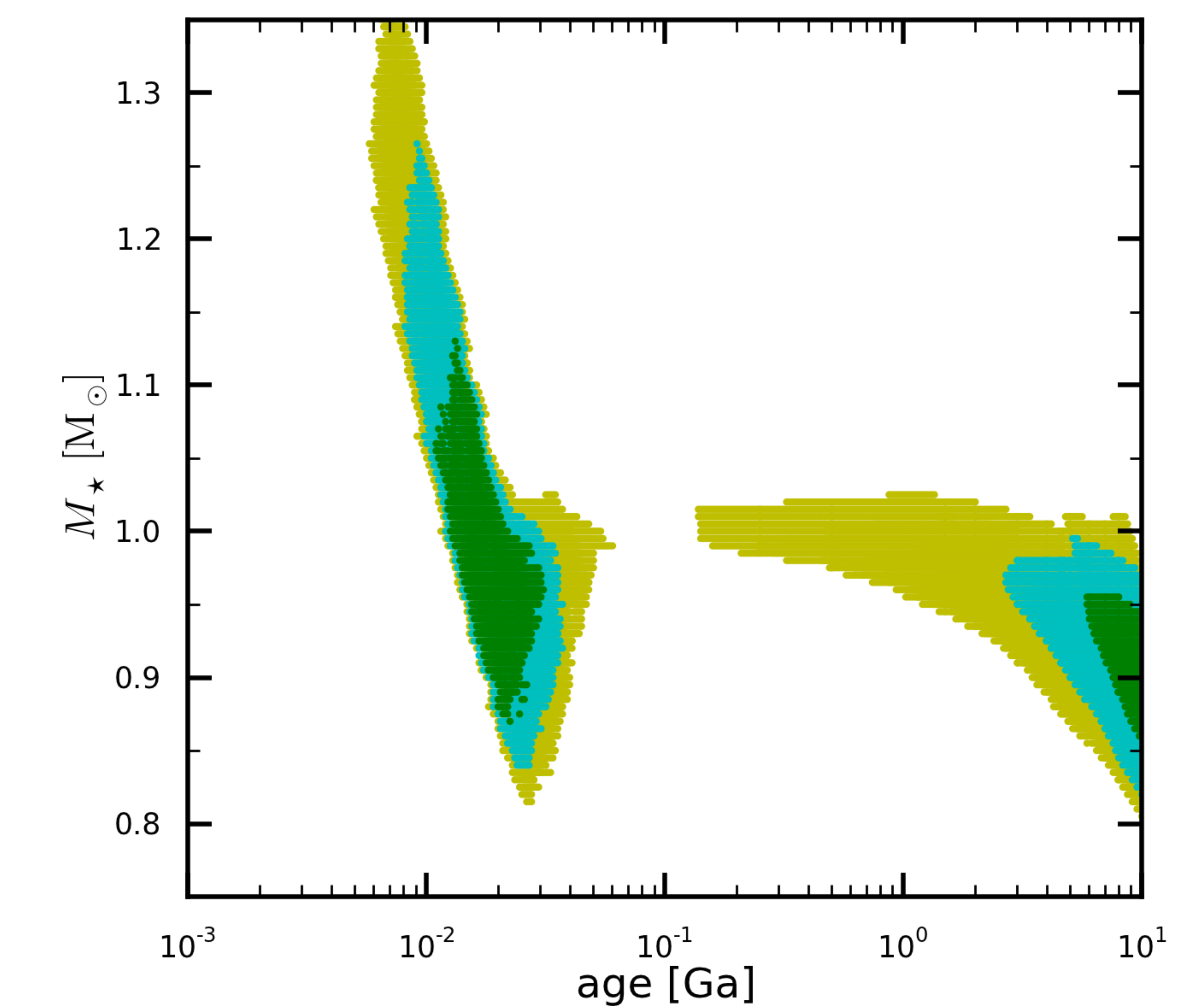}
\caption{Constraints obtained from stellar evolution models for the age and mass of  \cible. 
The colored circles correspond to constraints derived from stellar evolution models 
matching the stellar density and effective temperature within a certain number of standard 
deviations: less than 1$\,\sigma$ (green), 2$\,\sigma$  (blue), or 3$\,\sigma$~(yellow).
}
\label{fig_age_star}
\end{center}
\end{figure}

However, the inferred stellar densities for  \cible\ and CoRoT-2
($1.35\pm0.25$ vs. $1.814_{-0.045}^{+0.050}\rm\,g\,cm^{-3}$ -- see 
Gillon et al.~\cite{gillon10}) slightly differ. 
As a result, the effective temperature, metallicity, and density constraint for  \cible\ are consistent with evolution 
tracks for solar-mass stars that are either particularly young and still on the pre-main sequence or 
old and towards the end of the main-sequence evolution (see Figs.~2-4 from Guillot \& 
Havel~\cite{guillot11}). We used the CESAM evolution code (Morel \& Lebreton~\cite{morel08}) 
to calculate these evolution tracks. In Fig.~\ref{fig_age_star} we plot the solutions in the stellar-mass, 
age parameters, with colors that depend on their quadratic distance to the effective temperature-stellar 
density constraints. These point towards either 
a young age, less than 30~Ma (at $3\sigma$), or an old one, more than 8~Ga at $1\sigma$ and more 
than 4~Ga at $3\sigma$. The situation is thus different than for CoRoT-2, for which a continuum of 
young and late-type solutions was found (see Guillot \& Havel~\cite{guillot11}). For \cible, these 
solutions contradict with the other age indicators. 

First, as for CoRoT-2, the rapid rotation of the star points towards a young age. 
Bouvier~(\cite{bouvier07}) has compiled hundreds of rotational period 
measurements from photometric surveys of young open clusters with ages 
up to 625~Ma (Hyades), and used them to model the rotational evolution 
of stars in several mass bins. A range of models is needed to
reproduce the data at any given age, and constraints are scarce beyond
200~Ma for stars with masses between 0.8 and 1\,\Msun. However,
extrapolation of the models calibrated at earlier ages suggests that
such stars are not expected to retain rates $\sim 5$ times faster than
the Sun (which is the case for \cible) beyond ages of 500\,-\,600\,Ma.

Second, the lithium equivalent width favors the latter end of this age range. 
According to a recent compilation of Li depletion measurements, also from 
stars in young open clusters (Hillenbrand et al.~\cite{hillenbrand09}), 
the equivalent width of 40\,m\AA\ measured for \cible\ is typical of stars 
of this spectral type at the age of Ursa Majoris (500\,Ma) or the Hyades 
(625\,Ma), while typical equivalent widths for similar stars in M34 and 
M7 ($\sim 200$\,Ma) are about 100\,m\AA. This would thus instead 
suggest an age of several hundred Ma for \cible.

Thus, for now we are unable to estimate the age of \cible, even 
if it seems to be young. 
This illustrates the difficulty in determining the age of stars.
We adopt the stellar mass $M_\star=0.95\pm0.15$\,\Msun.  
The conservative error bar is large enough to agree with both pre-main-sequence case
and old star main-sequence star. 
This implies a semi-major axis 
$a = 0.0295 \pm 0.0016$, a stellar radius
$R_\star = 1.00 \pm0.13$\,\Rsun, and thus a planetary radius
$R_{\rm p} = 1.31\pm0.18$\,\Rjup. 
The equilibrium temperature of the planet assuming an isotropic zero-albedo 
is $T_{\rm eq}= 1550\pm90$\,K.
The rotation period $5.4 \pm 0.4$~days (Sect.~\ref{sect_rota_period})  
is consistent with the high value \vsini$\,=8.0\pm1.0$\,\kms\ and is
fast for a late G-type star. Using $R_\star=1.00\pm0.13$\,\Rsun, this translates 
into an inclination of the stellar rotation axis that is $i_* = 70^{\circ} \pm 20^{\circ}$,
so the star is seen nearly edge-on.

\subsection{Radial velocities fit}
\label{sect_RV_fit}

We fitted the radial velocities with a Keplerian model. The period and the epoch of 
the transit were fixed to the values obtained from the light curves analysis
(Sect.~\ref{sect_LC_analysis}).
If the relative accuracy of the radial velocity measurements is high (a few 
tens of \ms\ here in the case of \cible), their absolute accuracy in heliocentric or 
barycentric frames could be ten times less good, so a 
radial velocity shift was free to vary in the fit between the three datasets 
used for the orbit (SOPHIE, HARPS/HAM, and FIES), and we finally obtained one 
systemic radial velocity for each of the three instruments. 
The HARPS/EGGS data 
secured during a transit do not significantly constrain the orbit, and they are used 
below for the analysis  of the Rossiter-McLaughlin anomaly. We found the upper 
limit $e<0.08$ at 95\%\ confidence
for the eccentricity of the orbit and thus assumed a circular orbit, 
as usually is the case for hot jupiters. In the case of a slightly eccentric orbit, 
its orientation is not well constrained, with the longitude of the periastron
included in the range $-60^{\circ} < \omega < 100^{\circ}$.

The final fit of the radial velocities is plotted in Fig.~\ref{fig_orb_rv}.
The derived orbital parameters are reported in 
Table~\ref{starplanet_param_table}, together with error bars that were 
computed from  $\chi^2$  variations and Monte~Carlo experiments. 
The radial velocity variations present a semi-amplitude 
$K=590\pm14$~m\,s$^{-1}$, corresponding to a planet with a mass 
\mp~$  = 3.47 \pm 0.38$~\Mjup. This assumes $M_\star = 0.95\pm0.15$\,M$_\odot$ 
for the host star, which here is the main source of uncertainty~on~\mp.

The standard deviation of the residuals to the fit is $\sigma_{O-C}=41.0$~m\,s$^{-1}$
for the whole dataset ($35.9$, $44.4$, and $43.0$\,\ms\ for SOPHIE, HARPS/HAM,
and FIES, respectively). The reduced $\chi^2$ is 1.02 for the 22 radial velocities used 
in the fit. We do not detect any drift over the 106-day span of the radial velocity, 
with an upper limit of 200\,\ms\,a$^{-1}$ at 95\%\ confidence. We can thus exclude any 
extra planet in the system with a mass higher than 3\,\Mjup\ and a period shorter 
than~200~days.

\subsection{Planetary evolution}   
\label{sect_evol}   

We have seen that \cible\ and CoRoT-2 are similar for what concerns their stars. Their planets 
(\cibleb\ \textit{vs.} CoRoT-2b) are also strikingly similar, in terms of orbital periods 
(1.90 \textit{vs.} 1.74 days), masses (3.4 \textit{vs.} 3.7\,\Mjup), and equilibrium temperatures 
(1550 \textit{vs.} 1539\,K). The inferred planetary radii are $1.31\pm 0.18$\,\Rjup\ for
\cibleb\ and $1.466_{-0.044}^{+0.042}$\,\Rjup\ for CoRoT-2b (Gillon et al.~\cite{gillon10}).

\begin{figure}[h] 
\begin{center} 
\includegraphics[scale=0.55]{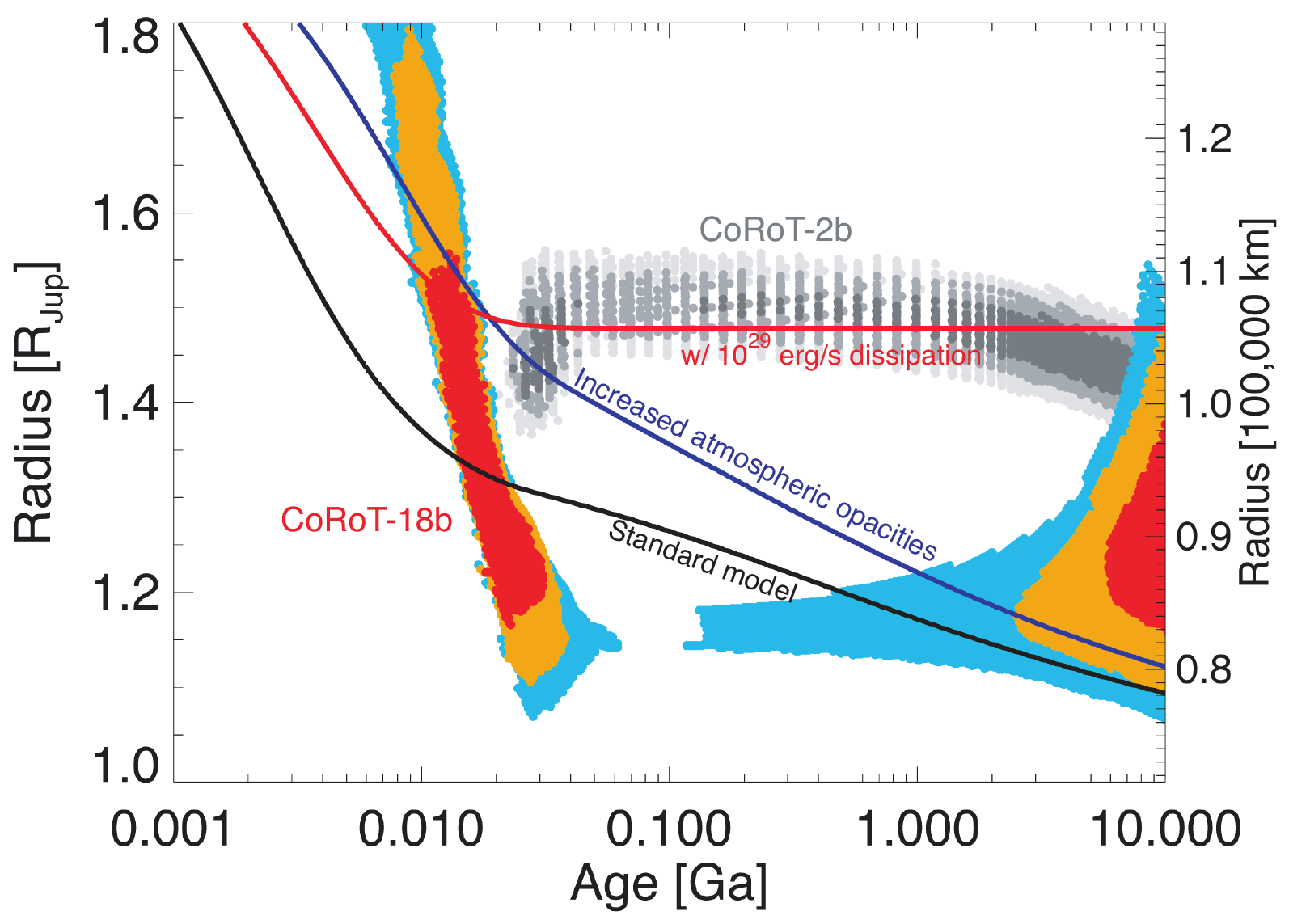}
\caption{Constraints obtained on the age and radius of  \cibleb. 
The colored circles correspond to 1\,$\sigma$ (red), 2\,$\sigma$ (yellow), or 3\,$\sigma$ (blue) 
solutions. The figure also plots the same constraints for CoRoT-2b (grayscale), when not 
including the effects of spots  (see Czesla et al.~\cite{zesla09}, 
Guillot \& Havel~\cite{guillot11}). The evolution tracks show the progressive 
contraction of a $3.5\,\rm M_{Jup}$ planet with $T_{\rm eq}\sim 1600$\,K, in the so-called 
``standard approach", when increasing atmospheric opacities by a factor 3.5, and when 
dissipating $10^{29}\rm\,erg\,s^{-1}$ at the center (see Guillot \& Havel~\cite{guillot11} 
for a description of the models). 
}
\label{fig_evolb}
\end{center}
\end{figure}

Figure~\ref{fig_evolb} compares the planetary radii obtained as a function of age for the two 
planets, using the approach described in Guillot \& Havel~(\cite{guillot11}). To use 
the same approach for the two planets, when calculating the size of CoRoT-2b, we did not 
account for the effect of spots (see Czesla et al.~\cite{zesla09}, 
Guillot \& Havel~\cite{guillot11}) -- in the case of CoRoT-2b, these yield a $\sim 3\%$ 
increase in the inferred radius. For CoRoT-2b, the age constraints from evolution models matching 
($T_{\rm eff}$, $\rho_*$, $\log g$, [Fe/H]) are weak and $3\sigma$ solutions are found anywhere 
between 20\,Ma to more than 15\,Ga). The inferred radius is extremely large for an object of this 
mass and planetary evolution models require a recent ($\sim 20\,$Ma) dramatic event 
(birth, giant impact, close encounter, and circularization of the orbit) to explain it. Because of its 
lower stellar density, solutions for \cible\ are found at both extremes of the age range (10 to 25\,Ma 
and more than 4\,Ga, see Sect.~\ref{sect_age}). 
The constraints on the size of \cible\ are weaker so that they cannot be used to confirm or refute 
possible inflation mechanisms or possible compositions, at least as long as the cause of the age 
mismatch between stellar evolution models and youth indicators (see Sect.~\ref{sect_age}) is not found. 

If the mismatch found for \cible\ between the various age indicators can be linked to the poorly 
known physics of young stars, it may lead to completely reconsider the problem of the inflated 
size of CoRoT-2b and other exoplanets orbiting young~stars.

\subsection{Rossiter-McLaughlin anomaly analysis}   
\label{sect_fit_RM}   

The radial velocity measured during the 2011 January 28 transit 
were fitted in order to derive the sky-projected angle $\lambda$ between 
the planetary orbital axis and the stellar rotation axis. 
We applied the analytical approach developed by Ohta et al.~(\cite{otha05})
to model the Rossiter-McLaughlin anomaly shape, which use here 
ten parameters: 
the stellar limb-darkening linear coefficient $\epsilon$, 
the transit parameters $R_\mathrm{p}/R_*$, $a/R_*$ and $i$, 
the parameters of the circular orbit ($P$, $T_0$, and $K$), 
the HARPS/EGGS systemic radial velocity,
and finally \vsini\ and $\lambda$.
We adopted $\epsilon=0.722$ computed by Claret~(\cite{claret04})
in the $g^\prime$ filter corresponding to the HARPS 
wavelength range. The transit and orbital parameters were determined above  
from the light curves and radial velocity fits, and   their uncertainties are negligible 
for the fit of the Rossiter-McLaughlin anomaly shape, according the uncertainties 
of the HARPS/EGGS radial velocities. The main parameters that play a role in 
this fit are the systemic velocity, $\lambda$, and \vsini. As these parameters 
are correlated in the Rossiter-McLaughlin fit, we computed the \kid\ of the fit on 
a three-dimensional grid scanning their possible values. 

\begin{figure}[h] 
\begin{center}
\includegraphics[scale=0.525]{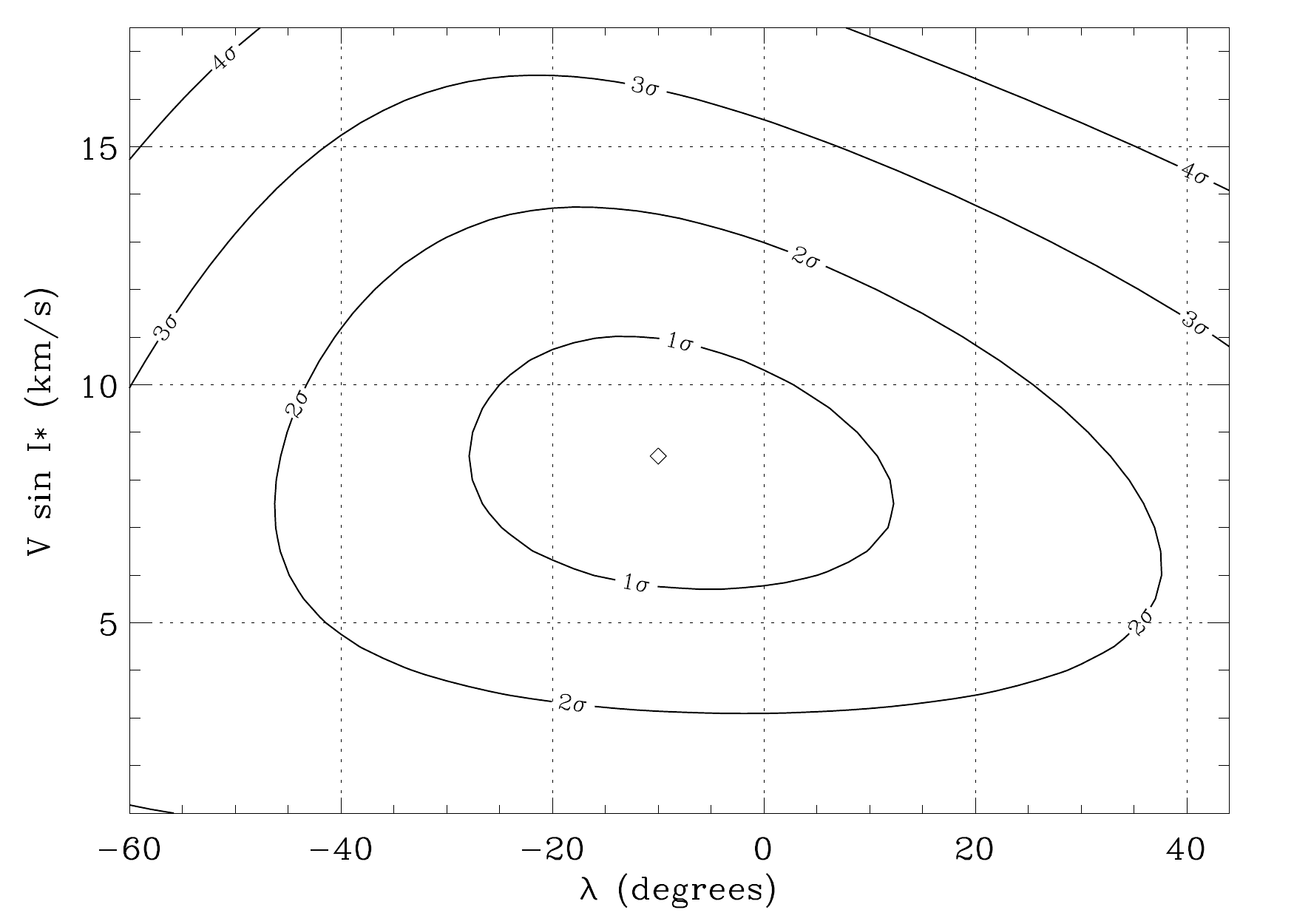}
\caption{\kid\ isocontours for our 
modeling of the Rossiter-McLaughlin effect as a function of $\lambda$ 
and  \vsini. The diamond shows the lowest-\kid\  value.
}
\label{fig_contour_RM}
\end{center}
\end{figure}

The systemic velocity we obtained for the transit observed with 
HARPS/EGGS,  $29.550\pm0.016 $, 
is similar to this obtained in Sect.~\ref{sect_RV_fit} 
for the orbit observed with HARPS/HAM.
It is constrained thanks to the observations secured immediately before and 
after the transit.
The confidence interval contours estimated from \kid\  variations
for the $\lambda$ and \vsini\ are plotted in Fig.~\ref{fig_contour_RM}.
We thus obtained $\lambda = -10^{\circ} \pm 20^{\circ}$ and 
\vsini$\, = 8.5 \pm 2.5$\,\kms. The best fit is plotted in Fig.~\ref{fig_RM}. 
The dispersion of the residuals to the fit is 46.5\,\ms. This agrees with the expected 
error bars on the radial velocities, and this is similar to the dispersion of the residuals 
to the fit of the orbit presented in Sect.~\ref{sect_RV_fit}. 

The Rossiter-McLaughlin anomaly  detection only stands on a few points; 
however, the shape of the radial velocity variations during the transit agree 
with a Rossiter-McLaughlin feature, 
with $\lambda \simeq0^{\circ}$ and the expected amplitude.
As a statistical test for the Rossiter-McLaughlin anomaly detection, 
we computed the \kid\
over the 12 measurements secured during the transit night, and 
we get 12.5 and 24.7 for the fits including or not the Rossiter-McLaughlin anomaly, 
respectively. Including the Rossiter-McLaughlin model in the fit thus implies a 
factor two improvement in the  \kid, for basically two extra free parameters 
($\lambda$ and \vsini, which mainly constraints the Rossiter-McLaughlin shape). 
We performed an F-test which indicates there is a probability $>70$\% 
that the \kid\ improvement~actually is due to the Rossiter-McLaughlin anomaly~detection.

Usually only the sky-projected value $\lambda$ of the obliquity $\psi$ could be 
measured because the inclination $i_*$ of the stellar rotation axis remains unknown. 
Here we estimated $i_* = 70^{\circ} \pm 20^{\circ}$ (Sect.~\ref{sect_spectral_analysis}), 
so the sky-projected $\lambda$ could be translated into the actual obliquity. 
We obtain $\psi = 20^{\circ} \pm 20^{\circ}$. This value remains inaccurate, due
to the significant uncertainties on $\lambda$ and $i_*$. It allows, however, 
the conclusion that the orbit of \cibleb\ is prograde and nearly aligned. 
This additionnaly reinforces  the similarity between the \cible\ and CoRoT-2
planetary systems,  since CoRoT-2b is
also prograde and aligned (Bouchy et al.~\cite{bouchy08}).

The \vsini\ value obtained 
from this fit agrees with this obtained in 
Sect.~\ref{sect_spectral_analysis} from spectral analysis, \vsini\,$=8.0\pm1.0$\,\kms.
As discussed by, e.g., Hirano et al.~(\cite{hirano10}) and Simpson 
et al.~(\cite{simpson10}), modeling the Rossiter-McLaughlin anomaly 
could produce biased  \vsini\ measurements when rotational broadening of the stellar 
lines is significantly larger than the instrument resolution. We do not see that effect here, 
possibly because of the long exposure times of each~exposure.

\begin{table*}
\begin{center}
\caption{Planet and star parameters.}            
\begin{tabular}{ll}        
\hline\hline                 
\\
\multicolumn{2}{l}{\emph{Fitted transit parameters}} \\
\hline
Planet orbital period $P$ [days] &  $1.9000693 \pm 0.0000028$  \\
Transit center epoch $T_0$ [HJD] & $2\,455\,321.72412\pm0.00018$  \\
Scaled semi-major axis $a/R_{*}$ & $6.35\pm0.40$ \\
Radius ratio $k=R_{\rm p}/R_{*}$ & $0.1341\pm0.0019$ \\
Impact parameter $b = a \cos i / R_{*} $ & $0.40^{+0.08}_{-0.14}$  \\
\\
\multicolumn{2}{l}{\emph{Deduced transit parameters}} \\
\hline
Orbit inclination $i$ [$^{\circ}$] & $86.5^{+1.4}_{-0.9}$  \\
Transit duration $T_{1-4}$ [h] & $2.387\pm0.037$ \\
Egress/ingress duration $T_{1-2} =T_{3-4}$ [h] & $0.331\pm0.038$ \\
$M^{1/3}_{*}/R_{*}$ [Solar units]& $0.99 \pm 0.06$ \\
Stellar density $\rho_{*}$ [$g\;cm^{-3}$] & $1.35\pm0.25$\\
\\
\multicolumn{2}{l}{\emph{Results from radial velocity observations}} \\
\hline    
Radial velocity semi-amplitude $K$ [\ms] & $590\pm14$ \\
Orbital eccentricity $e$  &  $<0.08$ \\
SOPHIE systemic velocity [\kms] & $29.533\pm0.016$ \\
HARPS/HAM systemic velocity [\kms] & $29.572\pm0.015$ \\
FIES  systemic velocity [\kms] & $29.657\pm0.034$ \\
O-C residuals [\ms] & 41.0\\
\\
\multicolumn{2}{l}{\emph{Spectroscopic stellar parameters }} \\
\hline
Effective temperature $T_{\rm eff}$[K] & $5440\pm100$ \\
Surface gravity log\,$g$ [dex]&  $4.4\pm0.1$  \\
Metallicity $[\rm{Fe/H}]$ [dex]&  $-0.1\pm0.1$ \\
Rotational velocity {\vsini} [\kms] & $8.0\pm1.0$ \\
Spectral type & G9V \\
Star mass [\Msun] &  $0.95\pm0.15$ \\
\\
\multicolumn{2}{l}{\emph{Stellar and planetary parameters}} \\
\hline
Star radius [\Rsun] &  $1.00\pm0.13$  \\
Distance of the system [pc] & $870 \pm90$ \\
Stellar rotation period $P_{\rm rot}$ [days]  &   $5.4\pm0.4$  \\
Stellar inclination $i_*$ [$^{\circ}$] &  $70 \pm 20$ \\
Orbital semi-major axis $a$ [AU] & $0.0295 \pm 0.0016 $ \\
Planet mass $M_{\rm p}$ [M$_J$ ] &   $3.47\pm0.38$ \\
Planet radius $R_{\rm p}$[R$_J$]  &  $1.31\pm0.18$ \\
Planet density $\rho_{\rm p}$ [$g\;cm^{-3}$] &  $2.2 \pm 0.8$ \\
Equilibrium temperature  $T_{\rm eq}$ [K] & $1550\pm90$ \\
\\
\multicolumn{2}{l}{\emph{Rossiter-McLaughlin parameters }} \\
\hline
Sky-projected obliquity $\lambda$ [$^{\circ}$] & $10\pm20$ \\
HARPS/EGGS systemic velocity [\kms] & $29.550\pm0.016 $ \\
O-C residuals (HARPS/EGGS) [\ms] & 46.5 \\
Obliquity $\psi$ [$^{\circ}$] & $20\pm20$ \\
\\
\hline       
\end{tabular}
\label{starplanet_param_table}  
\end{center}
\end{table*}

\section{Conclusion}
\label{sect_concle}

We reported the detection of the 18$^{\rm th}$ transiting exoplanet detected by the CoRoT project.
This giant planet was discovered thanks to the high-accuracy, continuous photometry obtained by the CoRoT 
satellite and the photometric and spectroscopic follow-up performed on ground-based telescopes. 
\cibleb\ is a massive hot jupiter orbiting a faint G9V star. Its mass is 
\mp\,$  = 3.47 \pm 0.38$\,\Mjup, and its radius 
$R_{\rm p} = 1.31\pm0.18$\,\Rjup, implying a density
$\rho_{\rm p} = 2.2 \pm 0.8$\,g/cm$^3$.
The period of the circular orbit is $1.9000693 \pm 0.0000028$~days. It is known with an accuracy 
better than 0.25~seconds and the mean mid-transit epoch with an accuracy of 20~seconds.
The mass of the host star  is $M_\star = 0.95\pm0.15$\,M$_\odot$ and its radius
$R_\star = 1.00 \pm0.13$\,R$_\odot$. The parameters of this system are summarized 
in Table~\ref{starplanet_param_table}. 

The parameters of \cibleb\ are similar to those of 
CoRoT-2b (Alonso~\cite{alonso08}; Gillon et al.~\cite{gillon10}), 
and to a lesser extent to those of 
CoRoT-11b (Gandolfi~\cite{gandolfi10}) and 
CoRoT-17b (Csizmadia et al.~\cite{csizmadia11}), 
the other massive hot jupiters found with CoRoT.
Interestingly \cibleb\ is found to be either particularly young (a few tens of Ma) 
or old ($>4$\,Ga) from stellar 
evolution models matching the star's effective temperature and inferred density, but according 
both to its lithium abundance and to its relatively fast rotation, it would be expected to be 
modestly young (several hundred Ma). 
This mismatch potentially points to a problem in our understanding of the evolution of 
young stars, with possibly significant implications for stellar physics and the interpretation of 
inferred sizes of exoplanets around young stars. 
In addition, \cibleb\ and CoRoT-2b are the only known planets (transiting or not) orbiting 
a fast-rotating, G-type or cooler star.

The orbit of \cibleb\ is prograde, with a spin-orbit angle  
$\psi = 20^{\circ} \pm 20^{\circ}$
(sky-projected value $\lambda = -10^{\circ} \pm 20^{\circ}$), 
hence an obliquity in agreement with 0.
Schlaufman~(\cite{schlaufman10}) and Winn et al.~(\cite{winn10b}) 
have shown that misaligned planets tend to orbit
hot stars. With an effective temperature of $5440 \pm 100$\,K  for the 
host star, the \cible\ system supports this trend. Exceptions to this trend 
are increasing, however (Moutou et al.~\cite{moutou11}; 
Brown et al.~\cite{brown11}).

\begin{figure}[h] 
\begin{center}
\includegraphics[scale=0.525]{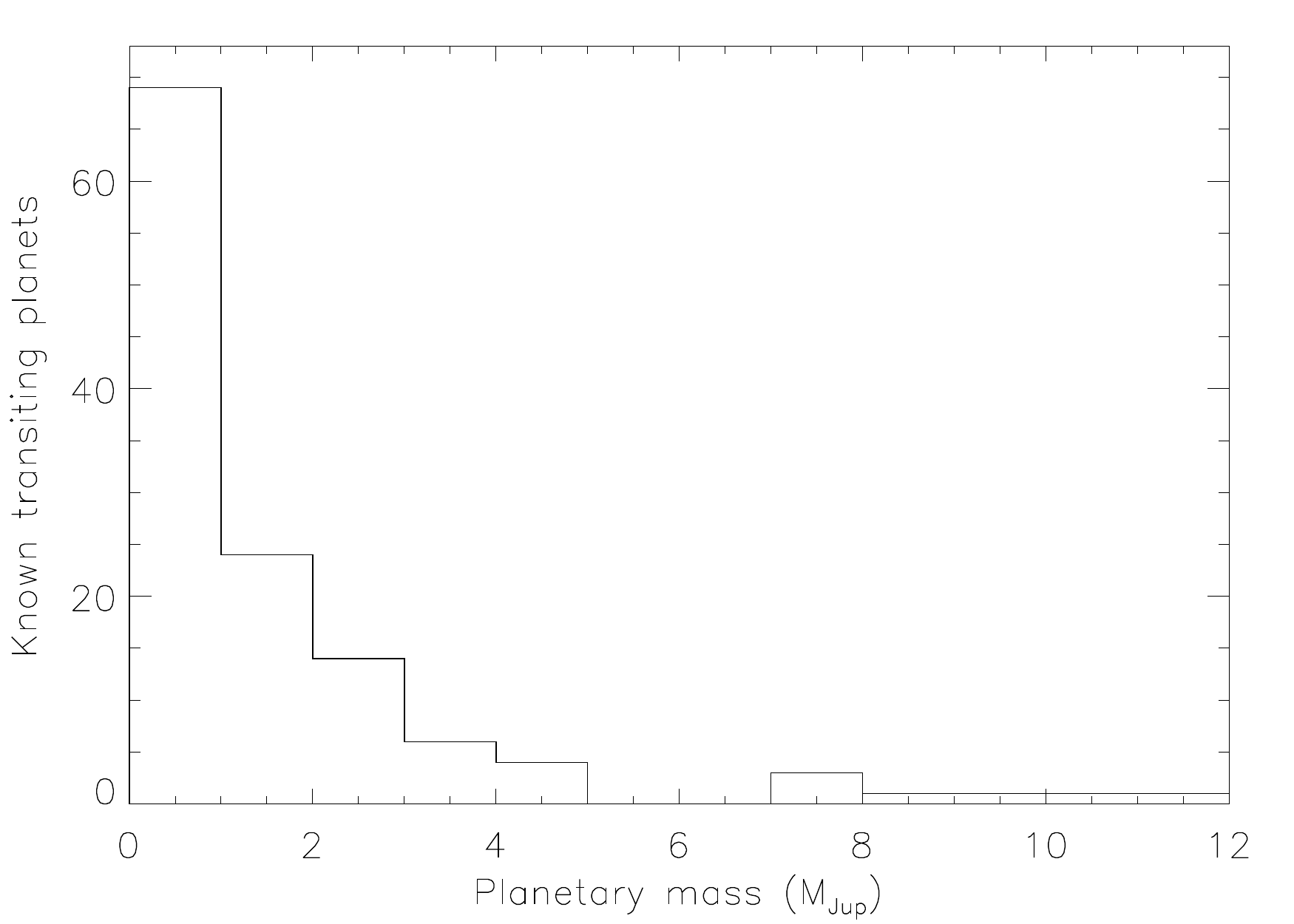}
\caption{Histogram of the number of known transiting planets as a function of their mass
(data from http://exoplanet.eu).
}
\label{fig_histo}
\end{center}
\end{figure}

H\'ebrard et al.~(\cite{hebrard10}; \cite{hebrard11}) have hypothesized 
that most of the massive planets are prograde and moderately but significantly 
misaligned, whereas the less massive planets are distributed in two thirds of the 
prograde, aligned systems and one third of the strongly misaligned systems.
Being prograde and nearly aligned, \cibleb\ is at the upper limit of the low-mass 
range. Interestingly, 
a limit is also apparent in the mass distribution of known transiting planets. 
This is shown in Fig.~\ref{fig_histo}, where one can see a decreasing 
abundance of planets with increasing planetary mass up to 4.5\,\Mjup.
No transiting planets are known in the range
\mp$  = [4.5-7]$\,\Mjup, and a few are known with \mp$>7$\,\Mjup. It is difficult 
to imagine that a bias is provoking the lack of massive planets, as they are easier 
to detect. This different distribution suggests that planets below  4.5\,\Mjup\ could 
have a different nature or history than those above 7\,\Mjup. This is reinforced 
by the different obliquity~distribution.

\begin{acknowledgements}
The French teams are grateful to the CNES for its constant support and the funding of AB, JMA, CC.
IAP/OHP team acknowledges support of French National Research Agency (ANR-08-JCJC-0102-01).
The team at the IAC acknowledges
support by grants ESP2007-65480-C02-02 and AYA2010-20982-C02-02 of the
Spanish Ministry of Science and Innovation (MICINN).
The German CoRoT Team (TLS and University of Cologne) acknowledges
DLR grants 50OW0204, 50OW0603, and 50QM1004.
We are grateful to N. Piskunov of the Uppsala Astronomical Observatory for continuing to make 
SME available to us and for answering questions about its implementation and operation.
SOPHIE observations (program 10B.PNP.MOUT) were done on the 1.93-m telescope at
Observatoire de Haute-Provence (CNRS), France.
HARPS observations (program 184.C-0639) were done on the 3.6-m telescope at the ESO
La Silla Paranal observatory, Chile.
FIES observations (program P42-216) were done on the Nordic Optical Telescope, operated 
on the island of La Palma jointly by Denmark, Finland, Iceland, Norway, and Sweden, in the 
Spanish Observatorio del Roque de los Muchachos of the Instituto de Astrofisica de Canarias.
\end{acknowledgements}

\end{document}